\documentclass[a4paper,11pt]{article}
\usepackage{jcappub}
\usepackage{graphicx}
\usepackage{amsmath}
\usepackage{amssymb}
\usepackage{hyperref}
\usepackage{physics}
\usepackage[capitalize]{cleveref}
\usepackage{subcaption}

\makeatletter
\gdef\@fpheader{\mbox{}}
\makeatother

\newlength{\fullw}
\newlength{\halfw}
\newlength{\threefigw}
\newlength{\twofigw}
\newlength{\onefigw}
\newlength{\bigfigw}
\newcommand\hcancel[2][black]{\setbox0=\hbox{$#2$}%
\rlap{\raisebox{.45\ht0}{\textcolor{#1}{\rule{\wd0}{0.5pt}}}}#2} 

\newcommand{\SUNDIALS}{\texttt{SUNDIALS}}
\newcommand{\FLINT}{\texttt{FLINT}}
\newcommand{\CUBA}{\texttt{CUBA}}
\newcommand{\Fortran}{\texttt{Fortran}}

\DeclareMathOperator{\rect}{rect}

\newcommand{\dirac}[1]{\delta\negthinspace\left(#1\right)}

\newcommand{\heaviside}[1]{\mathrm{\Theta}\!\left( #1 \right)}

\newcommand{\jtheta}[2]{\vartheta_{#1}\negthinspace\left(#2\right)}
\newcommand{\jthetab}[2]{\vartheta_{#1}\negthinspace\left[#2\right]}
\newcommand{\djtheta}[2]{\vartheta_{#1}^{\prime}\negthinspace\left(#2\right)}
\newcommand{\djthetab}[2]{\vartheta_{#1}^{\prime}\negthinspace\left[#2\right]}

\newcommand{\Eone}[1]{E_{1}\negthinspace\left(#1\right)}
\newcommand{\Eoneb}[1]{E_{1}\negthinspace\left[#1\right]}

\newcommand{\Evalb}[1]{\mathbb{E}\negthinspace\left[#1\right]}

\newcommand{\sss}[1]{{\scriptscriptstyle{#1}}}
\newcommand{\efolds}{$e$-folds}
\newcommand{\efold}{$e$-fold}

\newcommand{\uPl}{\mathrm{Pl}}

\newcommand{\uinf}{\mathrm{inf}}
\newcommand{\uqw}{\mathrm{qw}}
\newcommand{\ur}{\mathrm{r}}

\newcommand{\uLT}{\mathrm{LT}}
\newcommand{\uFPT}{\mathrm{FPT}}
\newcommand{\ufw}{\mathrm{fw}}
\newcommand{\usssPl}{\sss{\uPl}}

\newcommand{\usssLT}{\sss{\uLT}}
\newcommand{\usssFPT}{\sss{\uFPT}}
\newcommand{\umode}{\mathrm{mode}}
\newcommand{\uini}{\mathrm{ini}}

\newcommand{\zero}{{_0}}
\newcommand{\one}{{_1}}

\newcommand{\Mp}{M_\usssPl}

\newcommand{\Nqw}{N_\uqw}

\newcommand{\Nzero}{N_\zero}
\newcommand{\None}{N_\one}

\newcommand{\Nflat}{N_{\flat}}

\newcommand{\deltaN}{\var N}
\newcommand{\deltan}{\var n}

\newcommand{\rvp}[1]{\bar{#1}}
\newcommand{\rv}[1]{\Delta#1}

\newcommand{\Pname}{P}
\newcommand{\Pof}[1]{\Pname\negthinspace\left(#1\right)}
\newcommand{\Pinf}[1]{\Pname_\infty\negthinspace\left(#1\right)}
\newcommand{\Pfw}[1]{\Pof{#1}}
\newcommand{\Prv}[1]{\rvp{\Pname}\negthinspace\left(#1\right)}
\newcommand{\Plt}[1]{\Pname_{\usssLT}\negthinspace\left(#1\right)}
\newcommand{\Pfpt}[1]{\mathbb{P}_{\usssFPT}\negthinspace\left(#1\right)}
\newcommand{\FTof}[1]{\hat{#1}}
\newcommand{\PfwFT}[1]{\FTof{\Pname}\negthinspace\left(#1\right)}

\newcommand{\Frv}{\rvp{F}}

\newcommand{\Wrv}{\rvp{W}}

\newcommand{\qzero}{q_\zero}

\newcommand{\Hinf}{H_\uinf}

\newcommand{\phiqw}{\phi_\uqw}
\newcommand{\phir}{\phi_\ur}
\newcommand{\phizero}{\phi_\zero}
\newcommand{\phione}{\phi_\one}

\newcommand{\chizero}{\chi_\zero}
\newcommand{\chir}{\chi_\ur}

\newcommand{\chihat}{\hat{\chi}}
\newcommand{\chizerohat}{\chihat_\zero}
\newcommand{\chirhat}{\chihat_\ur}

\newcommand{\xzero}{x_\zero}
\newcommand{\wzero}{w_\uini}
\newcommand{\tzero}{t_\zero}

\newcommand{\calN}{\mathcal{N}}

\newcommand{\zetahat}{\hat{\zeta}}

\newcommand{\zetafw}{\zeta_{\ufw}}
\newcommand{\zetamode}{\zeta_{\umode}}
\newcommand{\zetamodeinf}{\zeta^{\infty}_{\umode}}

\begin{document}

\title{Time-reversed stochastic inflation in the quantum well}

\author[a]{Chiara Animali,}
\author[a]{Baptiste Blachier,}
\author[b]{Nanoka Okada,}
\author[a]{\\Christophe Ringeval,}
\author[c]{Tomo Takahashi,}
\author[d]{and Koki Tokeshi}

\affiliation[a]{\textit{%
    Cosmology, Universe and Relativity at Louvain (CURL),\\%
    Institute of Mathematics and Physics, University of Louvain,\\%
    2 Chemin du Cyclotron, 1348 Louvain-la-Neuve, Belgium%
}}%
\affiliation[b]{\textit{%
    Graduate School of Science and Engineering, Saga University, Saga 840-8502, Japan%
}}%
\affiliation[c]{\textit{%
    Department of Physics, Saga University, Saga 840-8502, Japan%
}}%
\affiliation[d]{\textit{%
    Laboratoire de Physique de l’\'{E}cole Normale Sup\'{e}rieure, ENS, CNRS,\\% 
    Universit\'{e} PSL, Sorbonne Universit\'{e}, Universit\'{e} Paris Cit\'{e}, 75005 Paris, France%
}}%

\emailAdd{chiara.animali@uclouvain.be}
\emailAdd{baptiste.blachier@uclouvain.be}
\emailAdd{nanoka.o20000925@gmail.com}
\emailAdd{christophe.ringeval@uclouvain.be}
\emailAdd{tomot@cc.saga-u.ac.jp}
\emailAdd{koki.tokeshi@phys.ens.fr}

\date{\today}

\abstract{Time-reversed stochastic inflation solves the stochastic
  evolution of the inflationary universe backward in time, by counting
  the number of $e$-folds from the end of quantum diffusion towards
  some initial state.  The point of view of observers attached to the
  end-of-inflation hypersurface is thus enforced. In this work, we
  exactly solve time-reversed stochastic inflation in a flat and
  bounded potential, the so-called quantum well. At given lifetime,
  the field behaviour is found to be either indistinguishable from the
  one obtained in a semi-infinite flat potential, or, subject to
  enhanced stochasticity where any memory of the initial state is
  erased.  The derived distribution of curvature perturbations reduces
  to the semi-inﬁnite result for small ﬂuctuations while it develops
  exponential tails for the large ones.  Such tails arise for both
  positive and negative values, and decay twice as fast as the one
  obtained in the standard “forward” stochastic inﬂation.  These
  differences may have important consequences for tail-sensitive
  phenomena, such as primordial black hole formation.}


\maketitle

\section{Introduction}
\label{sec:intro}

Cosmic Inflation has become the leading paradigm that describes the
earliest moments of the universe, offering a causal mechanism for the
origin of primordial fluctuations while resolving the horizon and
flatness issues of the standard hot Big-Bang
model~\cite{Starobinsky:1979ty, Starobinsky:1980te, Sato:1980yn,
  Guth:1980zm, Linde:1981mu, Albrecht:1982wi, Linde:1983gd,
  Mukhanov:1981xt, Mukhanov:1982nu, Starobinsky:1982ee, Guth:1982ec,
  Hawking:1982cz, Bardeen:1983qw}. The paradigm assumes that an epoch
of accelerated expansion of the spacetime has occurred before the
standard radiation-dominated era of the Friedmann-Lema\^{\i}tre
model. An inflationary era can be triggered by a scalar field having a
potential energy large enough to dominate the energy budget of the
universe~\cite{East:2015ggf, Linde:2017pwt, Aurrekoetxea:2019fhr,
  Joana:2020rxm, Joana:2022uwc}. In the semi-classical regime,
inflation is sustained while the field slowly rolls down a flat enough
potential. At the same time, quantum fluctuations in the field-metric
system are stretched to astrophysical length scales and can be shown
to generate nearly scale-invariant and approximately Gaussian
curvature fluctuations and primordial gravitational
waves~\cite{Mukhanov:1987pv, Mukhanov:1988jd, Mukhanov:1990me,
  Stewart:1993bc} (see also Refs.~\cite{Auclair:2022yxs,
  Bianchi:2024qyp} for recent advances).  Within inflationary
cosmology, these primordial fluctuations are the seeds of the density
perturbations observed in the cosmic microwave background and
large-scale structures~\cite{Planck:2018jri, Planck:2019kim,
  EIopiparous,Martin:2024qnn}.

The semi-classical description requires the quantum fluctuations to
remain subdominant compared to the classical field evolution driving
the accelerated expansion. However, it is possible for the quantum
fluctuations to dominate, either because the Hubble parameter during
inflation is large, typically of the same order as the reduced Planck
mass $\Mp$, or because the potential is very flat and the classical
evolution is strongly suppressed. In this regime, referred to as
``quantum diffusion'', the evolution of the universe can be described
by the language of stochastic processes~\cite{Starobinsky:1986fx,
  Goncharov:1987ir, Nambu:1987ef, Kandrup:1988sc, Nakao:1988yi,
  Starobinsky:1994bd, Linde:1993xx, Salopek:1998qh}. For models with a
single scalar field evolving in a nearly flat potential $V$, the
stochastic inflation formalism provides an effective field-theoretical
formulation for the infrared (IR) modes. Here, the IR sector is
defined with respect to the Hubble scale, and consists of modes whose
wavelengths are larger than the Hubble radius. Quantum fluctuations
are initially generated in the ultraviolet (UV) sector, namely on
subhorizon scales, and are subsequently stretched beyond the Hubble
radius by the exponential expansion, thereby becoming part of the IR
sector. In other words, the IR mode reservoir undergoes a continuous
and random inflow from the UV sector. One can then define a
coarse-grained field $\phi$ which describes the dynamics of the IR
modes only. Assuming slow-roll, it is found to satisfy a Langevin
equation
\begin{equation}
\dv{\phi}{N} = -\dfrac{1}{3 H^2} \dv{V}{\phi} + \dfrac{H}{2\pi} \xi(N),
\label{eq:langevin}
\end{equation}
in Planck units ($\Mp=1$). The quantity $H(\phi)$ stands for the
Hubble parameter during stochastic inflation, $N=\ln a$ is the number
of {\efolds}, $a$ being a scale factor for the
Friedmann-Lema\^{\i}tre-Robertson-Walker (FLRW) metric, and $\xi(N)$
is a Gaussian white noise coming from the collective effect of all UV
modes~\cite{Grain:2017dqa}. This equation appears homogeneous due to
the choice of a suitable coordinate system, the uniform $N$ gauge. The
time variable has to be $N$, the number of {\efolds}, to ensure
consistency with quantum field theory
expectations~\cite{Finelli:2008zg, Finelli:2010sh,
  Vennin:2015hra}. Let us stress that, by construction, the
stochasticity of \cref{eq:langevin} generated by the noise term
$\xi(N)$ is representative of the underlying quantum fluctuations of
the system. Although the stochastic inflation formalism washes out
quantum entanglement, it is a non-perturbative theory that preserves
quantum randomness~\cite{Vennin:2020kng}. As such, under some
assumptions, it is possible and justified not to distinguish
stochastic realisations of $\phi$ from quantum ones. In other words,
starting from a deterministic field value $\phizero$ at a given time
$N=\Nzero$, the non-vanishing noise $\xi(N)$ generates many different
solutions of \cref{eq:langevin}, whose distribution provides a
stochastic representation of their quantum realisations.

Since the advent of inflation it has been realised that quantum
diffusion can lead to everlasting inflationary realisations and strong
inhomogeneities~\cite{Vilenkin:1983xq, Linde:1986fd, Goncharov:1987ir,
  Winitzki:2006rn, Winitzki:2008yb, Creminelli:2008es,
  Tomberg:2025fku}. As can be seen in \cref{eq:langevin}, in an
exactly flat potential, the first term on the right hand side vanishes
and $\phi(N)$ would be a pure Brownian motion. Large fluctuations of
$\phi$ imply a strongly inhomogeneous spacetime. In fact, the
curvature fluctuations can be quantitatively determined from the
stochastic processes associated with \cref{eq:langevin} by using the
so-called stochastic-$\deltaN$ formalism~\cite{Fujita:2013cna,
  Fujita:2014tja, Ando:2020fjm, Mizuguchi:2024kbl, Launay:2024qsm}. It
is a generalisation of the semi-classical $\deltaN$ formalism which
relates curvature fluctuations on super-Hubble scales to variations in
the number of {\efolds}~\cite{Sasaki:1995aw, Sasaki:1998ug,
  Wands:2000dp, Lyth:2004gb, Lyth:2005fi}. For stochastic inflation,
the curvature fluctuations are given by $\zetafw= \calN - \Nflat$,
where $\calN$ is the elapsed number of {\efolds} during quantum
diffusion, a random variable. The quantity $\Nflat$ stands for a
reference unperturbed number of {\efolds}, usually set to be the
stochastic average $\Nflat =\ev{\calN}$. In the following, we will
refer to this formalism as ``forward'', as the {\efold} number
$\calN$ is counted forward in time. In this respect, it is closer to the so-called
$\deltan$ formalism of Ref.~\cite{Cruces:2025typ} than to the
standard $\deltaN$ formalism, in which {\efolds} are counted from the
end of inflation.

Most of the literature on stochastic inflation solves the inflationary
dynamics in the forward time direction, whereas observables are
defined on the end-of-inflation hypersurface, and this can lead to
difficulties. For instance, one can show that for a semi-infinite flat
potential, a shape typical of the plateau models of inflation, the
expectation value $\ev{\calN} =\infty$. As such the (forward)
curvature fluctuations $\zetafw$ are undefined, and usually assumed to
be infinite. Ref.~\cite{Blachier:2025tcq} introduced a time-reversed
approach to stochastic inflation that allows one to reverse the time in
the stochastic processes associated with \cref{eq:langevin}. The
number of {\efolds} is still stochastic but now counted from the end
of inflation towards the initial state while being conditioned by the
lifetime of the processes (the realisations of the random variable
$\calN$). The time-reversed formalism leads to well-defined
probability distributions for the reverse {\efold} numbers and their
associated curvature fluctuations $\zeta$. For the aforementioned
semi-infinite flat potential, one can show that the probability
distribution of $\Pof{\zeta}$ is normalisable, depends only on the
initial field value $\phizero$, and has tails decaying as
$1/\abs{\zeta}^{3/2}$. As such, this distribution does not have any
finite moments, but nothing particularly dangerous occurs for
semi-infinite flat potentials in spite of the divergences seen in the
forward approach. The divergences of the forward formalism disappear
when one considers a constant drift term in \cref{eq:langevin} (first
term in the right hand side). This drift would mimic the effects
associated with a tilted semi-infinite potential and this case has
been studied in Ref.~\cite{Blachier:2025iwk} in the time-reversed
formalism. Such a setup allows us to quantitatively compare the curvature
fluctuation distribution derived in the time-reversed and forward approaches to
stochastic inflation. Although they end up being qualitatively
similar, since both genuinely predict exponentially decaying tails for
the distribution of curvature perturbations, there are notable
differences. For instance, the forward distribution $P(\zetafw)$ is
only a one-sided exponential at $\zetafw>0$, while it behaves as a
bump function for negative values of $\zetafw$. On the contrary, the
reverse distribution $\Pof{\zeta}$ is exponential for both $\zeta>0$
and $\zeta<0$, but it decays exactly twice as fast as the positive
tails of $\Pof{\zetafw}$. Moreover, in the limit of infinite drift, in
which stochastic effects become negligible, only the time-reversed
picture recovers Gaussian tails for $\Pof{\zeta}$, whereas the forward
distribution remains a one-sided exponential. These differences are
not unexpected, since the time reversal enforces a local-observer
point of view with a uniform hypersurface at the end of quantum
diffusion~\cite{Tada:2016pmk}.
\begin{figure}
\begin{center}
  \includegraphics*[width=\onefigw]{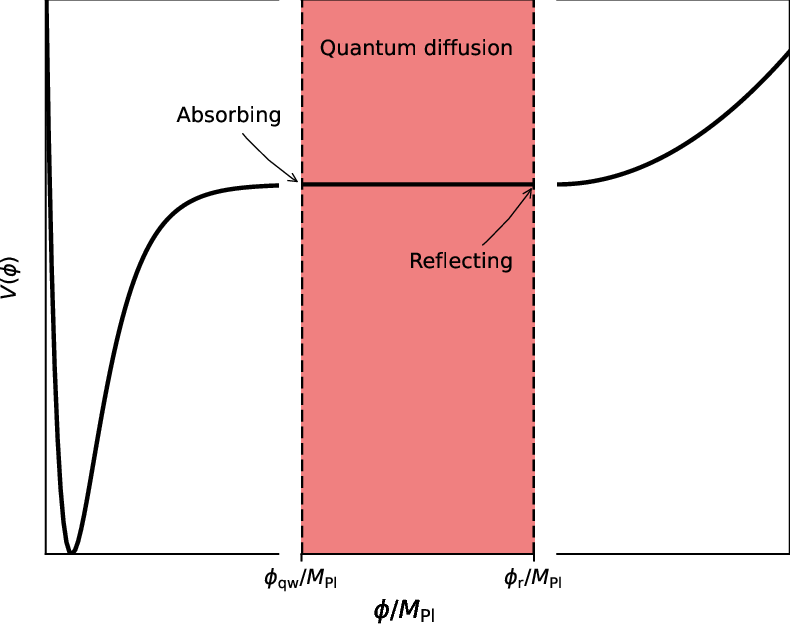}
  \caption{Sketch of the flat-well stochastic model. Quantum diffusion
    is considered on a flat and compact region of the potential, with
    an absorbing boundary condition at $\phi=\phiqw$ and a reflecting
    one at $\phi=\phir$. The boundary conditions ensure that quantum
    diffusion ends forever at $\phi<\phiqw$ while preventing it to
    explore the region $\phi>\phir$. An illustration of a possible
    consistent completion of the potential has also been represented.}
  \label{fig:flatwell}
\end{center}
\end{figure}

In this article, we solve time-reversed stochastic inflation in the
\emph{finite} flat-well model. The potential is assumed to be exactly
flat over a compact domain, as represented in
\cref{fig:flatwell}. Unlike the semi-infinite flat potential, the
flat-well setup is non-pathological in the forward approach and
provides another simple setting for comparing the distributions of
curvature perturbations derived from the two approaches. Moreover, a
flat well, also referred to as ``quantum well'', has been intensively
studied in the context of primordial black holes (PBHs). Indeed,
quantum diffusion generated by a small flat domain of the inflationary
potential located after the end of semi-classical inflation may
generate large curvature fluctuations susceptible to later collapse
into PBHs~\cite{Pattison:2017mbe, Ezquiaga:2019ftu, Ando:2020fjm,
  Vennin:2020kng, Pattison:2021oen, Tada:2021zzj, Tomberg:2022mkt,
  Animali:2022otk, Raatikainen:2023bzk, Stamou:2023vft,
  Stamou:2023vwz, Animali:2024jiz,Tomberg:2025fku, Animali:2025pyf,
  Raatikainen:2025gpd}.

As we show below, quantum diffusion in the flat well is quantitatively
the same as that occurring in the semi-infinite flat potential
whenever the field excursion across the well (the width) is larger
than the diffusion coefficient, i.e., for $\phir-\phiqw \gtrsim
H/(2\pi)$. In the opposite regime of a small well width, we find that
time-reversed stochastic inflation exhibits a greater amount of
stochasticity than in the semi-infinite flat potential, up to erasing
sensitivity to the initial field value $\phizero$. The probability
distribution of the curvature fluctuations is then derived and we
recover the same differences with respect to the forward approach as
those found in Ref.~\cite{Blachier:2025iwk}. The tails are
exponential, quasi-symmetric, and decay twice as fast as in the
forward picture.

The organisation of this article is as follows. A brief review of both
the forward and time-reversed formulations of stochastic inflation is
presented in \cref{sec:trinf}. In \cref{sec:fwmodel}, we focus on the
flat-well model and derive new exact solutions for the time-reversed
equations. In particular, we present two methods to calculate the
time-reversed transition probability distribution of the field values
and reverse {\efold} number, one of which makes explicit the relation
with the constrained stochastic processes discussed in
Ref.~\cite{Tokeshi:2023swe}. Based on these solutions, \cref{sec:curv}
is dedicated to the distribution of the curvature perturbation at
given lifetime, and to its marginalisation over all lifetimes. Our
results are then compared to the curvature fluctuation distribution derived from
the forward formalism. We give our conclusions in \cref{sec:concl}.

\section{Stochastic inflation and its time reversal}
\label{sec:trinf}

In this section, we review the general formalism used to solve for the
probability distribution of field values associated with
\cref{eq:langevin}, in both the standard, or forward, formulation
and in the time-reversed approach.

\subsection{Standard stochastic inflation}
\label{subsec:fwd}

The coarse-grained field $\phi$ is the solution of \cref{eq:langevin},
which is an It\^o stochastic differential equation of the generic form
\begin{equation}
\dd{\phi} = F\qty[\phi(N),N]\dd{N} + G\qty[\phi(N),N] \dd{W},
\label{eq:itofw}
\end{equation}
where $\dd W=\xi(N) \dd{N}$ is a Wiener process since we have
\begin{equation}
    \expval{ \xi (N) } = 0, \qquad \expval{ \xi (N_{1}) \xi (N_{2}) } = \dirac{N_{1} - N_{2}}.
\end{equation} 
The function $F=-V_{,\phi}/(3 H^2)$ is a friction term, also called a
drift, while the diffusion amplitude is encoded in $G=H/(2\pi)$. The
transition probability density for the solutions $\Bqty{\phi(N)}$ of
the It\^o process in \cref{eq:itofw} is solution of the Fokker-Planck
(or forward Kolmogorov) equation given by~\cite{Sarkka_Solin_2019}
\begin{equation}
\pdv{\Pfw{\phi,N\mid\phizero,\Nzero}}{N} =
-\pdv{}{\phi}\qty[F(\phi,N)\Pfw{\phi,N\mid\phizero,\Nzero}] +
\dfrac{1}{2}\pdv[2]{}{\phi}\qty[G^2(\phi,N) \Pfw{\phi,N\mid\phizero,\Nzero}],
\label{eq:fp}
\end{equation}
where $N \ge \Nzero$. Let us stress that, by virtue of
this linear and parabolic differential equation, the transition
probability distribution $\Pfw{\phi,N\mid\phizero,\Nzero}$ is also the
Green's function of \cref{eq:fp}, i.e., a solution satisfying
\begin{equation}
\Pfw{\phi,N=\Nzero\mid\phizero,\Nzero} = \dirac{\phi-\phizero}.
\end{equation}
However, a unique solution of \cref{eq:fp} can be determined only once
the boundary conditions have been specified. The transition
probability $\Pfw{\phi,N\mid\phizero,\Nzero}$ contains all the
relevant information for the stochastic process, as it gives the
probability of getting a field value $\phi$ at {\efold} number $N$
knowing a field value $\phizero$ at a previous time
$\Nzero$. Concerning the elapsed number of {\efolds} $\calN$, it is
related to the so-called survival probability, namely the probability
of remaining in the quantum diffusion domain at a given {\efold}
number $N$. Taking as an illustrative example the flat well of
\cref{fig:flatwell}, it reads~\cite{Ando:2020fjm}
\begin{equation}
S(N\mid\phizero,\Nzero) \equiv
\int_{\phiqw}^{\phir}\Pfw{\phi,N\mid\phizero,\Nzero}\dd{\phi} = 1 -
\int_{\Nzero}^{N}\Plt{\Nqw-\Nzero\mid\phizero} \dd{\Nqw},
\label{eq:survival}
\end{equation}
where the integration variable $\Nqw$ denotes the time at which the
field reaches the exit boundary at $\phi=\phiqw$. Differentiating by
$N$ and using \cref{eq:fp} gives $\Plt{\rv{\Nzero}\mid\phizero}$, the
probability distribution of the lifetimes
\begin{equation}
\rv{\Nzero} = \Nqw - \Nzero,
\label{eq:rvNzero}
\end{equation}
which are the realisations of the number of elapsed {\efolds}
$\calN$. Using $N$ as a dummy variable, one obtains
\begin{equation}
  \begin{aligned}
\Plt{N-\Nzero\mid\phizero} &= -\pdv{S(N\mid\phizero,\Nzero)}{N} \\ & =
\Bqty{F\qty[\phi(N),N]\Pfw{\phi,N\mid\phizero,\Nzero} -
\dfrac{1}{2}
\pdv{\qty[G^2(\phi,N)\Pfw{\phi,N\mid\phizero,\Nzero}]}{\phi}}_{\phir}
\\ & - \Bqty{F\qty[\phi(N),N]\Pfw{\phi,N\mid\phizero,\Nzero} -
\dfrac{1}{2}
\pdv{\qty[G^2(\phi,N)\Pfw{\phi,N\mid\phizero,\Nzero}]}{\phi}}_{\phiqw}. 
\label{eq:Pfpt}
\end{aligned}
\end{equation}

For a unique exit boundary, as considered here, the probability
distribution of the lifetimes is also the probability distribution of
the first passage times at $\phi=\phiqw$ and one has
\begin{equation}
\Pfpt{N=\rv{\Nzero}+\Nzero\mid\phizero,\Nzero} = \Plt{\rv{\Nzero}\mid\phizero}.
\label{eq:Plt}
\end{equation}
In more complex situations, both distributions may differ, as for
instance in the presence of multiple exit boundaries or in the existence of
other killing mechanisms to stop the stochastic evolution. As such, in
spite of their equality, we will use the notation $\Plt{}$ when
referring to lifetimes and $\Pfpt{}$ when referring to first passage
times in the rest of the paper. According to the (forward)
stochastic-$\deltaN$ formalism, and as already discussed in the
introduction, the distribution of the curvature perturbation is given
by
\begin{equation}
\Pof{\zetafw\mid\phizero,\Nzero} = \Plt{\rv{\Nzero} = \zetafw + \Nflat\mid \phizero},
\label{eq:Pzetafwdef}
\end{equation}
where
\begin{equation}
\Nflat \equiv \int_{0}^{\infty} \rv{\Nzero} \,
\Plt{\rv{\Nzero} \mid \phizero} \dd{\rv{\Nzero}}.
\label{eq:Nflatdef}
\end{equation}

\subsection{Time-reversed stochastic inflation}
\label{subsec:rev}

Time-reversing stochastic inflation consists in considering the
stochastic process of \cref{eq:itofw} reversed in time. For instance,
in the potential sketched in \cref{fig:flatwell}, the stochastic field
$\phi$ emerges at $\phi=\phiqw$ to randomly evolve towards a sink
located at $\phi=\phizero$. The starting times have to be the ending
times of the forward processes, i.e., the first passage times $\Nqw$
at which each realisation of \cref{eq:itofw} reaches the absorbing
boundary at $\phi=\phiqw$. Their distribution is precisely given by
\cref{eq:Pfpt}. Markov processes, to which the process under scrutiny
belongs, can always be time-reversed, the past and future states being
independent given the present state of the
system~\cite{chung2005markov}. 

\subsubsection{Time-reversed Fokker-Planck equation}
As shown in
Refs.~\cite{Nagasawa1964,Anderson1982,Blachier:2025tcq}, it is
possible to derive a so-called reverse Fokker-Planck
equation\footnote{Not to be confused with the so-called \emph{backward}
Kolmogorov equation that is different.}  satisfied by the transition probability
$\Prv{\phi,\rv{N}\mid\phizero,\rv{\Nzero}}$ of the reverse process
\begin{equation}
  \begin{aligned}
\pdv{\Prv{\phi,\rv{N}\mid\phizero,\rv{\Nzero}}}{\rv{N}} =
& -\pdv{\phi}\left[\Frv(\phi,\rv{N})
  \Prv{\phi,\rv{N}\mid\phizero,\rv{\Nzero}}\right] \\ & + \dfrac{1}{2}
\pdv[2]{}{\phi}\qty[G^2(\phi,\rv{N})\Prv{\phi,\rv{N}\mid\phizero,\rv{\Nzero}}].
\end{aligned}
\label{eq:rvfp}
\end{equation}
The quantity $\rv{N}$ refers to the reverse {\efold} number, defined
by
\begin{equation}
\rv{N} \equiv \Nqw - N,
\label{eq:rvN}
\end{equation}
where $\Nqw$ is the {\efold} number from which the time-reversal is
performed. For our purpose this will be the time at which the field,
in the forward description, crosses the exit boundary at
$\phi=\phiqw$. $\Delta N$ ranges from $\rv{N}=0$, when the
reverse process starts, to $\rv{\Nzero}$ given by
\cref{eq:rvNzero}, at which the process hits $\phizero$. As such, the reverse
transition probability has to verify the following initial and
boundary conditions
\begin{equation}
\Prv{\phi,\rv{N}=0\mid\phizero,\rv{\Nzero}} = \dirac{\rv{\phi}},\quad
\Prv{\phi,\rv{N}=\rv{\Nzero}\mid\phizero,\rv{\Nzero}} = \dirac{\rv{\phi}-\rv{\phizero}},
\label{eq:Prvibc}
\end{equation}
where we have defined
\begin{equation}
\rv{\phi} \equiv \phi-\phiqw, \qquad \rv{\phizero} \equiv \phizero - \phiqw.
\label{eq:rvphi}
\end{equation}
The reversed drift term appearing in \cref{eq:rvfp} reads~\cite{Blachier:2025tcq}
\begin{equation}
    \Frv(\phi,\rv{N}) = -F(\phi,N) +
    \dfrac{1}{\Pfw{\phi,N\mid\phizero,\Nzero}} \pdv{}{\phi}\left[G^2(\phi,N)\Pfw{\phi,N\mid\phizero,\Nzero} \right],
\label{eq:rvdrift}
\end{equation}
which involves both the forward friction $F(\phi,N)$ and the
forward transition probability $\Pfw{\phi,N\mid\phizero,\Nzero}$. The
diffusion coefficient $G$ is exactly the same as in \cref{eq:fp}, a
Wiener process is indeed invariant under a time reversal.\footnote{Let 
us mention that the reverse Fokker-Planck equation~\eqref{eq:rvfp}
is stemming from an It\^o stochastic differential equation describing
the reverse stochastic process
\begin{equation}
\dd{\phi} = \Frv\qty[\phi(\rv{N}),\rv{N}]\dd{\rv{N}} + G\qty[\phi(\rv{N}),\rv{N}] \dd{\Wrv},
\label{eq:itorev}
\end{equation}
where $\Frv$ is given in \cref{eq:rvdrift} and $\Wrv$ is a Wiener
process.}

It is important to notice that the reverse transition probability
$\Prv{\phi,\rv{N}\mid\phizero,\rv{\Nzero}}$, solution of
\cref{eq:rvfp,eq:Prvibc}, is conditioned by both $\phizero$ and
$\rv{\Nzero}$, which therefore act as parameters. Indeed, a time
reversal from $(\phiqw,\Nqw)$ implies that we select random
realisations of \cref{eq:itofw} that necessarily cross the exit
boundary with unity probability. Equally, these very same trajectories
will reach $(\phizero,\Nzero)$ exactly at a reverse {\efold} number
$\rv{N} = \rv{\Nzero}$, both conditions being explicit in
\cref{eq:Prvibc}.

Among all possible realisations of the forward processes, the
time-reversal procedure consists of partitioning them into sub-ensembles
containing realisations that all have the same lifetime
$\rv{\Nzero}$. Within each sub-ensemble, all statistical properties are
given by the distribution $\Prv{\phi,\rv{N}\mid\phizero,\rv{\Nzero}}$, thereby
explaining the conditioning on $\rv{\Nzero}$. In order to recover
probability distributions over the whole ensemble, one should, at the
end of the day, reunite all the sub-ensembles weighted by their
respective probability distribution, i.e., the one of the lifetimes
$\Plt{\rv{\Nzero}\mid\phizero}$ given in \cref{eq:Plt}.

\subsubsection{Stochastic-$\delta N$ formalism in the time-reversed picture}

Another key point is that, at given lifetime $\rv{\Nzero}$, there exist, at every fixed 
field value $\rv{\phi}$, fluctuations in the reverse {\efold} number $\rv{N}$.
This is not a specificity of the time-reversed formalism itself.
Indeed, for any generic stochastic trajectory, a \textit{local time}
-- also called occupation time -- can be formally defined, which is a measure of the amount 
of time that the trajectory spends at a given field value, see e.g.~\cite{karlin1981second, Green01051986, Borodin2002}. 
Since it depends on each sample
path, it stands as a stochastic variable. In our case, the ensemble of such
local times defines a stochastic process for $\rv{N}$ endowed with the probability distribution
$\Pof{\rv{N}\mid\phi,\phizero,\rv{\Nzero}}$.

As a consequence, the stochastic-$\deltaN$ formalism can be framed at
given lifetime $\rv{\Nzero}$. Explicitly, from $\zeta
\equiv N-\Nzero - \ev{N-\Nzero}$, using \cref{eq:rvN,eq:rvNzero}, one gets
\begin{equation}
  \zeta = \ev{\rv{N}} - \rv{N},
\label{eq:zetadef}
\end{equation}
where the term $\ev{\rv{\Nzero}} = \rv{\Nzero}$ cancels out. The
lifetime is indeed not fluctuating, thanks to the conditioning of the
time-reversed formalism. It is important to notice that the expectation
value appearing in \cref{eq:zetadef} is over all the time-reversed
realisations, at given lifetime $\rv{\Nzero}$, i.e., it is the first moment
of the local times distribution $\Pof{\rv{N}\mid\phi,\phizero,\rv{\Nzero}}$. This is different from
\cref{eq:Pzetafwdef}, which postulates that $\zetafw$ is generated by
the fluctuations of the lifetime
itself~\cite{Blachier:2025tcq, Blachier:2025iwk}. From
\cref{eq:zetadef}, the time-reversed transition probability
distribution $\Prv{\phi,\rv{N}\mid\phizero,\rv{\Nzero}}$ allows us to
define a joint probability 
\begin{equation}
\Pof{\phi,\zeta\mid\phizero,\rv{\Nzero}} \propto \Prv{\phi,\rv{N}=\ev{\rv{N}}-\zeta\mid\phizero,\rv{\Nzero}}.
\label{eq:Pofzetaphidef}
\end{equation}
It requires the determination of
$\ev{\rv{N}}$ as a function of $\phi$, $\phizero$ and
$\rv{\Nzero}$, which is a technically difficult task. We will also further comment on its 
normalisation in \cref{sec:zetahat}. Once $\ev{\rv{N}}$ is computed, it is then possible to marginalise over
all field realisations to obtain
\begin{equation}
\Pof{\zeta\mid\phizero,\rv{\Nzero}} = \int
\Pof{\phi,\zeta\mid\phizero,\rv{\Nzero}} \dd{\phi},
\label{eq:Pofzetahatdef}
\end{equation}
which gives the probability distribution of $\zeta$, still at given
lifetime $\rv{\Nzero}$. Because, as observers attached to the
end-of-inflation hypersurface, we do not know in which of these
realisations we are, the final probability distribution of $\zeta$ has
to be marginalised over all possible lifetimes and one finally gets
\begin{equation}
\Pof{\zeta\mid\phizero} = \int_{0}^{\infty}
\Pof{\zeta\mid\phizero,\rv{\Nzero}} \Plt{\rv{\Nzero}\mid\phizero} \dd{\rv{\Nzero}}.
\label{eq:Pofzetadef}
\end{equation}

Let us now apply the time-reversed formalism to the finite flat-well
model presented in \cref{sec:intro}.

\section{Finite flat-well potential}
\label{sec:fwmodel}

In this section, we specialise to the flat-well setup represented in
\cref{fig:flatwell} and derive an exact solution for time-reversed
stochastic inflation. As discussed in \cref{subsec:rev}, the drift term
in the reverse Fokker-Planck equation~\eqref{eq:rvfp} takes as an
input the forward transition probability, namely, the solution of
\cref{eq:fp}. Thus, one needs first the solution of the forward
problem, which has been, in the context of inflation, originally
derived in Ref.~\cite{Ando:2020fjm}.

\subsection{Forward solution}
\label{sec:fwdsol}
Within the flat well, the potential $V(\phi)$ is constant and the
drift term $F$ of \cref{eq:fp} vanishes. From the first
Friedmann-Lema\^{\i}tre equation, the Hubble parameter during inflation
is constant
\begin{equation}
H^2(\phi) \simeq \dfrac{V(\phi)}{3 \Mp^2} \equiv \Hinf^2,
\end{equation}
and so it is for the diffusion coefficient
\begin{equation}
G(\phi,N) = G \equiv \dfrac{\Hinf}{2 \pi}.
\end{equation}
One has to solve a simple linear parabolic equation
\begin{equation}
\pdv{\Pfw{\phi,N\mid\phizero,\Nzero}}{N} = \dfrac{G^2}{2} \pdv[2]{\Pfw{\phi,N\mid\phizero,\Nzero}}{\phi},
\label{eq:basicheat}
\end{equation}
subject to the initial and boundary conditions
\begin{equation}
\Pfw{\phi,\Nzero\mid\phizero,\Nzero} = \dirac{\phi-\phizero}, \qquad
\Pfw{\phiqw,N\mid\phizero,\Nzero} = 0, \qquad
\eval{\pdv{\Pfw{\phi,N\mid\phizero,\Nzero}}{\phi}}_{\phir} = 0.
\label{eq:fwdbic}
\end{equation}
This equation can be solved in different manners, as for instance by
using a Fourier transform. The derivation has been detailed in
Appendix~\ref{sec:fwdfourier} and the solution is expressed as an
infinite sum over trigonometric functions that can be further
simplified to~\cite{Ando:2020fjm}
\begin{equation}
\Pfw{\phi,N\mid\phizero,\Nzero} =
\dfrac{1}{2\rv{\phir}} \qty[ \jtheta{2}{\dfrac{\pi}{2}
    \dfrac{\rv{\phizero-\rv{\phi}}}{\rv{\phir}},q} -  \jtheta{2}{\dfrac{\pi}{2}
    \dfrac{\rv{\phizero+\rv{\phi}}}{\rv{\phir}},q} ],
\label{eq:Pfwtheta}
\end{equation}
with
\begin{equation}
  q(N) = \exp[-\frac{\pi^2 G^2}{2\rv{\phir}^2}\left(N-\Nzero\right)],
  \qquad \rv{\phir} \equiv \phir - \phiqw.
\end{equation}
The second Jacobi theta function appearing in \cref{eq:Pfwtheta} is
defined by
\begin{equation}
\jtheta{2}{z,q} \equiv 2 \sum_{n=0}^{\infty} q^{\qty(n+1/2)^2} \cos\qty[(2n+1)z].
\label{eq:deftheta2}
\end{equation}
Note also that the quantity $\rv{\phir}$ naturally corresponds to the
width of the quantum well. From \cref{eq:Pfpt}, the probability of the
first passage times is readily obtained from \cref{eq:Pfwtheta} and
reads~\cite{Pattison:2017mbe}
\begin{equation}
  \Pfpt{N\mid\phizero,\Nzero} = \dfrac{G^2}{2} \eval{\pdv{\qty[\Pfw{\phi,N\mid\phizero,\Nzero}]}{\phi}}_{\phiqw} = -\dfrac{\pi}{4} \dfrac{G^2}{\rv{\phir}^2}
\djtheta{2}{\dfrac{\pi}{2} \dfrac{\rv{\phizero}}{\rv{\phir}},q},
\label{eq:fwPfpt}
\end{equation}
where $\djtheta{2}{z,q}$ stands for the derivative of
$\jtheta{2}{z,q}$ with respect to $z$.

Let us remark that $G \times \Pfw{\phi,N\mid\phizero,\Nzero}$ and
$\Pfpt{N\mid\phizero,\Nzero}$ only depend on the rescaled and
dimensionless field values
\begin{equation}
x \equiv \dfrac{\rv{\phi}}{\rv{\phir}}, \qquad \xzero =
\dfrac{\rv{\phizero}}{\rv{\phir}}, \qquad \chir \equiv \dfrac{\rv{\phir}}{G}\,.
\label{eq:xchidef}
\end{equation}
The quantity $x\in[0,1]$ is the field value, in reference to the
absorbing boundary, in units of the flat-well width $\rv{\phir}$. The
dimensionless quantity $\chir > 0$ is the width of the well, measured
in units of the diffusion coefficient and can be smaller or greater than unity.

\subsection{Time-reversed transition probability distribution}

We now turn to the time-reversed problem. The reverse drift term
appearing in \cref{eq:rvfp} is given by \cref{eq:rvdrift}, with
$F(\phi,N)=0$ for the flat well. Using \cref{eq:Pfwtheta}, it reads
\begin{equation}
\Frv\qty(\phi,\rv{N}) =G^2
\pdv{\ln \Pfw{\phi,N\mid\phizero,\Nzero}}{\phi}=
- \dfrac{\pi}{2} \dfrac{G}{\chir}
\dfrac{\djtheta{2}{\pi\, \dfrac{\xzero-x}{2},q} + \djtheta{2}{\pi\,\dfrac{\xzero+x}{2},q}}{\jtheta{2}{\pi\, \dfrac{\xzero-x}{2},q} - \jtheta{2}{\pi\,\dfrac{\xzero+x}{2},q}}\,,
\label{eq:Frvfpfw}
\end{equation}
with
\begin{equation}
q(\rv{N}) = \exp \qty[ 
    -\frac{\pi^2}{2\chir^2} \qty(\rv{\Nzero}-\rv{N}) 
    ] .
\end{equation}

The reverse transition probability distribution
$\Prv{\phi,\rv{N}\mid\phizero,\rv{\Nzero}}$, in the flat well, is the
unique solution of
\begin{equation}
  \begin{aligned}
\pdv{\Prv{\phi,\rv{N}\mid\phizero,\rv{\Nzero}}}{\rv{N}} & =
\dfrac{\pi}{2} \dfrac{G}{\chir} \pdv{}{\phi}\Bqty{
\dfrac{\djtheta{2}{\pi\, \dfrac{\xzero-x}{2},q} +
  \djtheta{2}{\pi\,\dfrac{\xzero+x}{2},q}}{\jtheta{2}{\pi\,
    \dfrac{\xzero-x}{2},q} -
  \jtheta{2}{\pi\,\dfrac{\xzero+x}{2},q}} \,
\Prv{\phi,\rv{N}\mid\phizero,\rv{\Nzero}} }
\\ &+
\dfrac{G^2}{2} \pdv[2]{\Prv{\phi,\rv{N}\mid\phizero,\rv{\Nzero}}}{\phi}\,,
\label{eq:rvfpfw}
\end{aligned}
\end{equation}
satisfying the boundary conditions in \cref{eq:Prvibc}.

\subsubsection{Solving the reverse Fokker-Planck equation}
\label{sec:solbyfp}

Although the drift term may seem ominous, it is possible to solve
\cref{eq:rvfpfw} exactly by virtue of the Maruyama-Girsanov's
theorem~\cite{Maruyama1955ContinuousMP, Girsanov1960,
  shreve2004stochastic, Sarkka_Solin_2019}. The details of the
derivation are presented in Appendix~\ref{sec:revmaruyama} and we
simply quote the solution here. It reads,
\begin{equation}
\Prv{\phi,\rv{N}\mid\phizero,\rv{\Nzero}} = \dfrac{1}{2 G \chir}
\dfrac{\djtheta{2}{\dfrac{\pi}{2}
    x,\qzero^{\tau}}}{\djtheta{2}{\dfrac{\pi}{2} \xzero,\qzero}}
\qty[\jtheta{2}{\pi \dfrac{\xzero-x}{2},\qzero^{1-\tau}} - \jtheta{2}{\pi \dfrac{\xzero+x}{2},\qzero^{1-\tau}}],
\label{eq:Prvsol}
\end{equation}
where we have defined
\begin{equation}
\qzero \equiv \exp \qty( -\frac{\pi^2}{2\chir^2}\rv{\Nzero} ), \qquad \tau
\equiv \dfrac{\rv{N}}{\rv{\Nzero}}\,,
\label{eq:qzerodef}
\end{equation}
both quantities encoding the dependence on the lifetime
$\rv{\Nzero}$. The rescaled time $\tau\in[0,1]$ measures the number of
reverse {\efolds} $\rv{N}$ in units of the lifetime $\rv{\Nzero}$.

One can explicitly check that \cref{eq:Prvsol} satisfies the required
initial and boundary conditions of \cref{eq:Prvibc} thereby ensuring
its unicity. Indeed, taking the limit $\rv{N}\to 0$, i.e., $\tau \to
0$, in \cref{eq:Prvsol} one has $\qzero^\tau \to 1^{-}$. One can then
use the scaling properties of the theta functions with respect to the
lattice parameter. From \cref{eq:deftheta2}, with $q=e^{-\pi t}$, one has~\cite{Thompson2011}
\begin{equation}
\sqrt{t} \, \jtheta{2}{z,e^{-\pi t}} = \exp \qty( -\frac{z^2}{\pi t} ) \jtheta{4}{\dfrac{i z}{t},e^{- \pi / t}},
\end{equation}
which allows us to derive their asymptotic form. For $q \to 1^{-}$, i.e., $t \to 0^+$,
one obtains
\begin{equation}
    \jtheta{2}{z,e^{-\pi t}} 
    \simeq 
    \dfrac{1}{\sqrt{t}} \, \exp \qty( -\frac{z^2}{\pi t} ), 
      \qquad 
      \djtheta{2}{z,e^{-\pi t}} \simeq
-\dfrac{2 z}{\pi t^{3/2}} \exp \qty( -\frac{z^2}{\pi t} ),
\label{eq:thetaqone}
\end{equation}
where we have used $\jtheta{4}{z,0}=1$. Using \cref{eq:thetaqone} in
\cref{eq:Prvsol}, in the limit $\rv{N}\to 0$, one gets
\begin{align}
    &\quad 
    \lim_{\rv{N} \to 0} \Prv{\phi,\rv{N}\mid\phizero,\rv{\Nzero}} 
    \notag \\ 
    &=
    -\lim_{\rv{N}\to0} \dfrac{x \sqrt{2}\chir^2}{G
      \qty(\pi\rv{N})^{3/2}} \dfrac{\jtheta{2}{\pi \dfrac{\xzero-x}{2},\qzero} -
  \jtheta{2}{\pi
    \dfrac{\xzero+x}{2},\qzero}}{\djtheta{2}{\dfrac{\pi}{2}
    \xzero,\qzero}}
    \exp \qty( -\frac{x^2\chir^2}{2\rv{N}} ) 
    \,.
\end{align}
For all $\phi\ne\phiqw$, i.e., $x\ne 0$, the theta-function ratio in this expression is finite
and the exponential factor ensures that
$\Prv{\phi,\rv{N}=0\mid\phizero,\rv{\Nzero}} = 0$. Taking the limit $x\to
0$, the numerator of the theta-function ratio approaches $-\pi x \djtheta{2}{\pi\xzero/2,\qzero}$ and
\begin{equation}
\lim_{\rv{N} \to 0} \Prv{\phiqw,\rv{N}\mid\phizero,\rv{\Nzero}} =
\lim_{\rv{N} \to 0} \sqrt{\dfrac{2}{\pi}} \dfrac{\rv{\phi}^2}{\qty(G^2\rv{N})^{3/2}} \exp(-\dfrac{\rv{\phi}^2}{2 G^2 \rv{N}}) = \dirac{\rv{\phi}}.
\end{equation}

Similarly, considering the limit $\rv{N}\to\rv{\Nzero}$,
i.e., $\tau\to1$, we have $\qzero^{1-\tau} \to 1^{-}$ in the argument
of the two theta functions of \cref{eq:Prvsol}. Using
\cref{eq:thetaqone}, one gets
\begin{equation}
\begin{aligned}
    &\quad 
    \lim_{\rv{N} \to \rv{\Nzero}} \Prv{\phi,\rv{N}\mid\phizero,\rv{\Nzero}} 
     \\ 
    &= \dfrac{\djtheta{2}{\dfrac{\pi}{2}
    \dfrac{\rv{\phi}}{\rv{\phir}},\qzero}}{\djtheta{2}{\dfrac{\pi}{2}
    \dfrac{\rv{\phizero}}{\rv{\phir}},\qzero}} % & \times
\lim_{\rv{N} \to \rv{\Nzero}}  \dfrac{ \exp \qty[ -\frac{\qty(\rv{\phi}-\rv{\phizero})^2}{2 G^2 \qty(\rv{\Nzero}
      - \rv{N})} ] - \exp \qty[ -\frac{\qty(\rv{\phi}+\rv{\phizero})^2}{2 G^2
      \qty(\rv{\Nzero}- \rv{N})} ]}{{\sqrt{2 \pi} \,
  G\sqrt{\rv{\Nzero}-\rv{N}}}}\,.
\end{aligned}
\end{equation}
The factor multiplying the theta-derivative ratio is the difference of two
Gaussian kernels and in the $\rv{N}\to\rv{\Nzero}$ limit is exactly the Dirac distribution
$\dirac{\rv{\phi}-\rv{\phizero}}$. The theta-derivative ratio is instead a function of
$\rv{\phi}$ that evaluates to unity for $\rv{\phi}=\rv{\phizero}$. As
such, the whole expression is also the distribution
$\dirac{\rv{\phi}-\rv{\phizero}}$, as expected from the boundary
condition.

\subsubsection{Inverting the forward process}

Another method to find the solution of \cref{eq:rvfpfw} is to use the
definition of the transition probability of a time-reversed process
from the state $(\phione,\None)$. For all $\Nzero<N<\None$, it
reads~\cite{nagasawa1993schroedinger, Nagasawa1964, Anderson1982,
  chung2005markov, Blachier:2025tcq}
\begin{equation}
\Prv{\phi,N\mid\phione,\None;\phizero,\Nzero} \equiv \dfrac{\Pfw{\phione,\None;\phi,N\mid\phizero,\Nzero}}{\Pfw{\phione,\None\mid\phizero,\Nzero}}\,.
\label{eq:Pfwdef}
\end{equation}
The joint probability can be expanded using the product rule as
\begin{equation}
\Pfw{\phione,\None;\phi,N\mid\phizero,\Nzero} =
\Pfw{\phi,N\mid\phizero,\Nzero} \Pfw{\phione,\None\mid\phi,N;\phizero,\Nzero}.
\label{eq:Pjoint:exp}
\end{equation}
The first factor in this expression is the transition probability of the
forward process to go from the initial condition $(\phizero,\Nzero)$
to the state of interest $(\phi,N)$. The second factor is also rooted in
the forward process and is the probability to jump from the state of
interest $(\phi,N)$ to a future state $(\phione,\None)$ given that the
process started at $(\phizero,\Nzero)$. For \emph{Markovian
processes}, which is the case here, and for $N>\Nzero$, one has
\begin{equation}
\Pfw{\phione,\None\mid\phi,N;\phizero,\Nzero} =
\Pfw{\phione,\None\mid\phi,N},
\label{eq:P:Mark}
\end{equation}
and the conditioning on the initial condition disappears. For the flat
well we are considering, the time reversal is performed
from the
absorbing boundary, i.e., $(\phione,\None)=(\phiqw,\Nqw)$. This may
appear problematic at first, since the denominator in \cref{eq:Pfwdef}
is indeed vanishing, $\Pfw{\phiqw,\Nqw\mid\phizero,\Nzero}=0$. However,
because $\Nqw$ is a random time, related to the lifetime of the
process, it can be shown that the limit $(\phione,\None) \to
(\phiqw,\Nqw)$ must be regular~\cite{chung2005markov}. Indeed, using
\cref{eq:Pjoint:exp} and \cref{eq:P:Mark} in \cref{eq:Pfwdef} and
taking the following limit
\begin{equation}
\Prv{\phi,\rv{N}\mid\phizero,\rv{\Nzero}} =
\Pfw{\phi,N\mid\phizero,\Nzero}  \lim_{\epsilon \to 0^+}
\dfrac{\Pfw{\phiqw+\epsilon,\Nqw\mid\phi,N}}{\Pfw{\phiqw+\epsilon,\Nqw\mid\phizero,\Nzero}},
\label{eq:infonnull}
\end{equation}
where
\begin{equation}
\Pfw{\phiqw+\epsilon,\Nqw\mid\phi,N} = \Pfw{\phiqw,\Nqw\mid\phi,N} +
\epsilon \eval{\pdv{\Pfw{\phione,\None\mid\phi,N}}{\phione}}_{\phione=\phiqw,\None=\Nqw} + \order{\epsilon^2},
\end{equation}
and an equivalent expression for the denominator of
\cref{eq:infonnull}, one gets
\begin{equation}
\Prv{\phi,\rv{N}\mid\phizero,\rv{\Nzero}} = \Pfw{\phi,N\mid\phizero,\Nzero}
\eval{\dfrac{\displaystyle\pdv{\Pfw{\phione,\None\mid\phi,N}}{\phione}}
  {\displaystyle\pdv{\Pfw{\phione,\None\mid\phizero,\Nzero}}{\phione}}}_{\phione=\phiqw,\None=\Nqw}.
\label{eq:Prvbydef}
\end{equation}
Plugging the expression of the forward transition probability written
in \cref{eq:Pfwtheta} into \cref{eq:Prvbydef} gives back the solution
presented in \cref{eq:Prvsol}.

\subsubsection{Relation with first passage times}

In another context Ref.~\cite{Tokeshi:2023swe} has considered
stochastic inflation processes constrained to realise a given number
of {\efolds}. It is intuitively expected that if that number of
{\efolds} is set to be the lifetime of the process, the probability
distribution for these constrained processes should be equivalent to
the one obtained from a time reversal on the unique exit boundary. In
the flat well, there is indeed no other way for the processes to stop
other than reaching $\phi=\phiqw$.

The generic expression for the first passage time probability distribution is given in
\cref{eq:Pfpt}. As explicit in \cref{eq:fwPfpt}, for vanishing drift
$F(\phi,N)=0$, a constant diffusion coefficient $G$ and a vanishing
first derivative at the reflective boundary $\phir$, one has the relation
\begin{equation}
\eval{\pdv{\Pfw{\phi,N\mid\phizero,\Nzero}}{\phi}}_{\phiqw} = \dfrac{2}{G^2}\Pfpt{N\mid\phizero,\Nzero}.
\end{equation}
Therefore, one may also rewrite \cref{eq:Prvbydef} as
\begin{equation}
\Prv{\phi,\rv{N}\mid\phizero,\rv{\Nzero}} = \Pfw{\phi,N\mid\phizero,\Nzero} \dfrac{\Pfpt{\Nqw\mid\phi,N}}{\Pfpt{\Nqw\mid\phizero,\Nzero}}\,.
\label{eq:Prvbyfpt}
\end{equation}
Let us stress that the above equality is not always valid. The
time-reversed transition probability distribution, the left-hand side
of this equation, is indeed defined by \cref{eq:Pfwdef} whereas the
right hand side is the definition of a stochastic process constrained
by its first passage times. Plugging \cref{eq:fwPfpt} into \cref{eq:Prvbyfpt} gives back
\cref{eq:Prvsol}.

\subsection{Recovering the semi-infinite potential}
\label{sec:semiinf}

\begin{figure}
\begin{center}
  \includegraphics*[width=\onefigw]{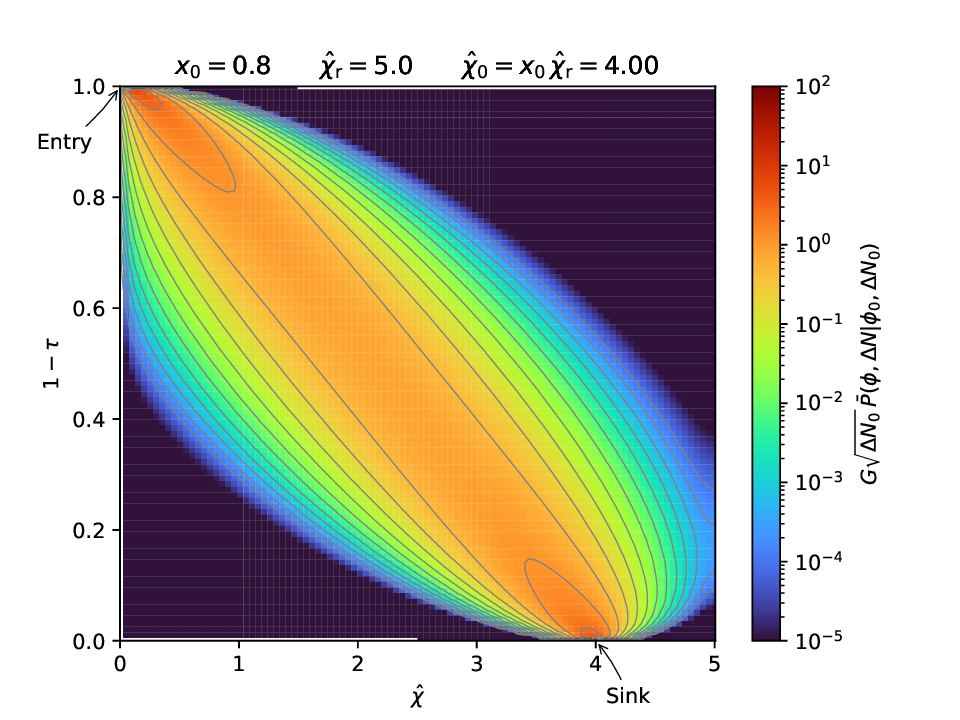}
\caption{Contour plots of the rescaled time-reversed transition
  probability distribution $G\sqrt{\rv{\Nzero}} \times
  \Prv{\phi,\rv{N}\mid\phizero,\rv{\Nzero}}$ given in \cref{eq:Prvsolhat} for a flat well having a
  width five times larger than the typical Brownian excursion
  ($\chirhat=5$). The quantity $\tau = \rv{N}/\rv{\Nzero}$ is the
  reverse {\efold} number in units of the lifetime $\rv{\Nzero}$. The
  horizontal axis $\chihat$ is the field value, in reference to the
  entry boundary, in units of $G\sqrt{\rv{\Nzero}}$, the typical
  Brownian excursion. The field emerges from the entry boundary, at
  $\chihat=0$ and $\tau=0$, and randomly evolves towards the sink
  located at $\chizerohat=4$ (the initial condition of the forward
  process), which is reached at $\tau=1$. For such a (large) value of
  $\chirhat$, this distribution is nearly identical to the one that
  would be obtained in the semi-infinite flat potential. The only visible
  differences here are the rightmost contours, in the tail of the
  distribution, not joining the sink but ending on the reflective
  boundary at $\chihat=\chirhat=5$.}
\label{fig:Prvlargew}
\end{center}
\end{figure}

Let us introduce another set of dimensionless field values, denoted $\chihat$, which are in reference to the boundary $\phiqw$, and
expressed in units of $G\sqrt{\rv{\Nzero}}$, the typical Brownian
excursion achieved during the lifetime $\rv{\Nzero}$. We define
\begin{equation}
\chihat \equiv \dfrac{\phi-\phiqw}{G \sqrt{\rv{\Nzero}}} =
\dfrac{\rv{\phi}}{G \sqrt{\rv{\Nzero}}} \equiv \dfrac{\chi}{\sqrt{\rv{\Nzero}}}\,,
\label{eq:chihatdef}
\end{equation}
and equivalent quantities for all the other field values, namely
\begin{equation}
\chirhat \equiv \dfrac{\phir-\phiqw}{G \sqrt{\rv{\Nzero}}} =
\dfrac{\rv{\phir}}{G \sqrt{\rv{\Nzero}}} \equiv
\dfrac{\chir}{\sqrt{\rv{\Nzero}}}\,, \qquad \chizerohat \equiv
\dfrac{\phizero-\phiqw}{G \sqrt{\rv{\Nzero}}} =
\dfrac{\rv{\phizero}}{G \sqrt{\rv{\Nzero}}} \equiv
\dfrac{\chizero}{\sqrt{\rv{\Nzero}}}\,.
\label{eq:chirhatdef}
\end{equation}
Let us remark that, from \cref{eq:xchidef}, the field values expressed in
units of the well width verify
\begin{equation}
x = \dfrac{\chi}{\chir} = \dfrac{\chihat}{\chirhat}, \qquad \xzero = \dfrac{\chizero}{\chir} = \dfrac{\chizerohat}{\chirhat}\,,
\label{eq:xotherdef}
\end{equation}
and one can rewrite \cref{eq:Prvsol} in terms of field values measured
in units of Brownian excursion. Using \cref{eq:chihatdef,eq:chirhatdef},
one obtains
\begin{equation}
  \begin{aligned}
G \sqrt{\rv{\Nzero}} \, \Prv{\phi,\rv{N}\mid\phizero,\rv{\Nzero}} & = \dfrac{1}{2\chirhat}
\dfrac{\djtheta{2}{\dfrac{\pi}{2}
    \dfrac{\chihat}{\chirhat},\qzero^\tau}}{\djtheta{2}{\dfrac{\pi}{2}
    \dfrac{\chizerohat}{\chirhat},\qzero}} \\ & \times 
\qty[\jtheta{2}{\dfrac{\pi}{2}
    \dfrac{\chizerohat-\chihat}{\chirhat},\qzero^{1-\tau}} -
  \jtheta{2}{\dfrac{\pi}{2}
    \dfrac{\chizerohat+\chihat}{\chirhat},\qzero^{1-\tau}}],
  \end{aligned}
  \label{eq:Prvsolhat}
\end{equation}
where, from \cref{eq:qzerodef}, we have
\begin{equation}
\qzero = \exp(-\dfrac{\pi^2}{2 \chirhat^2}).
\label{eq:qzerodefhat}
\end{equation}
Let us stress that, in terms of ``hat'' quantities, the rescaled
distribution, $\Prv{\phi,\rv{N}\mid\phizero,\rv{\Nzero}}$ multiplied by
$G\sqrt{\rv{\Nzero}}$, is universal as it no longer has any explicit
dependence on the lifetimes $\rv{\Nzero}$.

Let us now consider the limit $\chirhat \gg 1$ in \cref{eq:Prvsolhat},
i.e, the well width $\rv{\phir}$ is much larger than the
typical Brownian excursion $G\sqrt{\rv{\Nzero}}$. From \cref{eq:qzerodefhat},
this limit implies that $\qzero \to 1^{-}$ and all the theta functions
and derivatives appearing in \cref{eq:Prvsolhat} can be expanded
according to \cref{eq:thetaqone}, in a way similar to what has been
done in \cref{sec:solbyfp} for checking the initial and boundary
conditions. We have
\begin{subequations}
    \begin{align}
    \jtheta{2}{\pi\dfrac{\xzero\pm x}{2},\qzero^{1-\tau}} 
    &\simeq
    \sqrt{\dfrac{2}{\pi}} 
    \dfrac{\chirhat}{\sqrt{1-\tau}}
    \exp \qty[ 
        -\frac{\qty(\chizerohat\pm\chihat)^2}{2\qty(1-\tau)} 
    ], % \qquad
    \\ 
    \dfrac{\djtheta{2}{\dfrac{\pi}{2}x,\qzero^{\tau}}}{\djtheta{2}{\dfrac{\pi}{2}\xzero,\qzero}}
    &\simeq 
    \dfrac{\chihat}{\tau^{3/2}\chizerohat}
    \exp \qty( 
        -\frac{\chihat^2}{2\tau} + \frac{\chizerohat^2}{2} 
    ) .
    \end{align}
\end{subequations}
Plugging these expressions into \cref{eq:Prvsolhat} gives
\begin{equation}
  \begin{aligned}
G \sqrt{\rv{\Nzero}} \, \Prv{\phi,\rv{N}\mid\phizero,\rv{\Nzero}}  & \simeq
\dfrac{\sqrt{2/\pi}}{\tau^{3/2} \sqrt{1-\tau}} \,
\dfrac{\chihat}{\chizerohat}
\sinh\left(\dfrac{\chihat\chizerohat}{1-\tau}\right)
e^{-\frac{\chihat^2 + \tau^2 \chizerohat^2}{2 \tau(1-\tau)}} \\ &
\equiv G \sqrt{\rv{\Nzero}}\,\Pinf{\phi,\rv{N}\mid\phizero,\rv{\Nzero}},
\label{eq:Prvinf}
\end{aligned}
  \end{equation}
where $\Pinf{\phi,\rv{N}\mid\phizero,\rv{\Nzero}}$ is the transition
probability distribution derived in
Refs.~\cite{Blachier:2025tcq,Blachier:2025iwk} for the semi-infinite
flat potential. As a result, all physical quantities derived for the
flat well should asymptotically approach their corresponding analogues
in the semi-infinite flat potential once $\chirhat$ is sufficiently
large.

In \cref{fig:Prvlargew}, we have plotted the exact probability
distribution $\Prv{\phi,\rv{N}\mid\phizero,\rv{\Nzero}}$ for the flat
well, multiplied by $G\sqrt{\rv{\Nzero}}$, whose expression is given
in \cref{eq:Prvsolhat}, as a function of the dimensionless field value
$\chihat$ and forward time $1-\tau$, for a width set at $\chirhat =
5$. The sink, the initial field value of the forward process, is
located at $\xzero=0.8$, i.e., at $\chizerohat = 4$. This distribution
is almost indistinguishable from the one that would be obtained by
using instead the semi-infinite limit of \cref{eq:Prvinf}. Given the
lifetime $\rv{\Nzero}$ of the reverse processes, $\chirhat=5$ means
that the width of the well $\rv{\phir}$ is five times larger than the
typically expected Brownian excursion $G\sqrt{\rv{\Nzero}}$. As
\cref{fig:Prvlargew} shows, the process is somehow free to diffuse
within the well and the effects coming from the reflective boundary
are very small: only the two rightmost contours of the distribution
are ending on the reflective wall, with a vanishing derivative,
instead of connecting the entry boundary and the sink.

\subsection{Small width and saturated quantum diffusion}

\label{sec:saturated}

\begin{figure}
\begin{center}
  \includegraphics*[width=\onefigw]{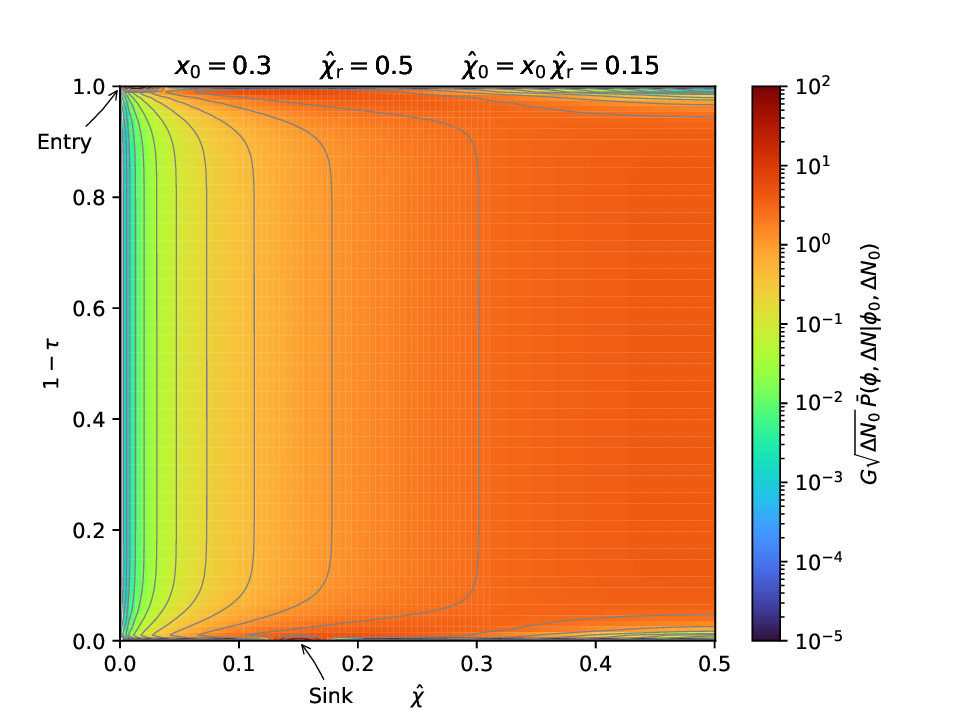}
  \includegraphics*[width=\onefigw]{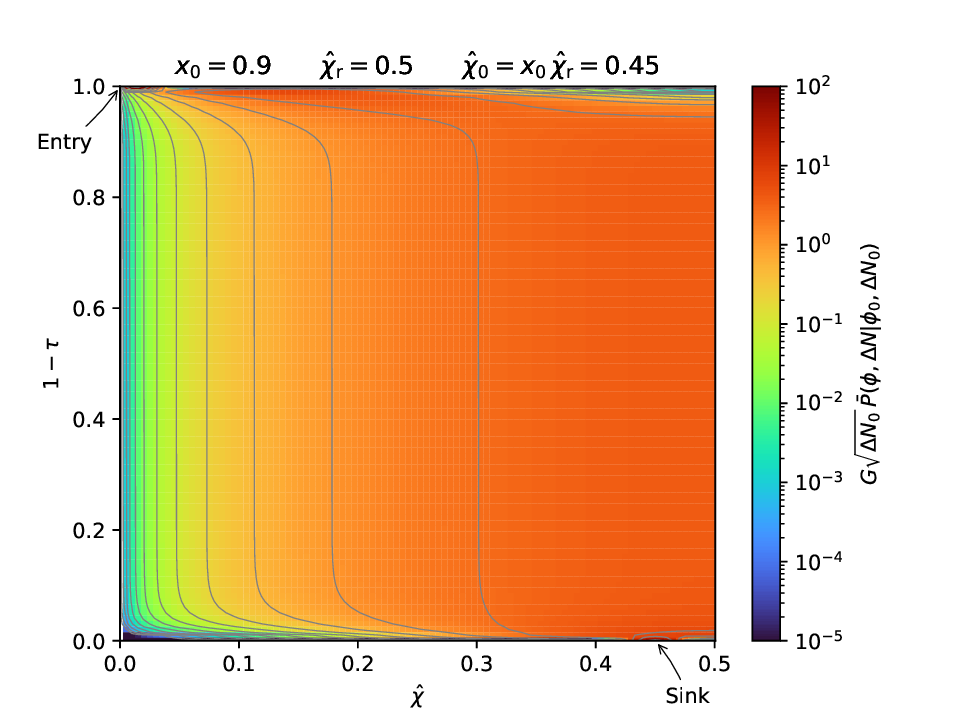}
\caption{Contour plots of the rescaled time-reversed transition
  probability distribution $G\sqrt{\rv{\Nzero}} \times
  \Prv{\phi,\rv{N}\mid\phizero,\rv{\Nzero}}$ given in \cref{eq:Prvsolhat} for a flat well having a
  width $\chirhat=0.5$, i.e., half of the typical Brownian excursion
  $G\sqrt{\rv{\Nzero}}$ (this is ten times smaller than the one of
  \cref{fig:Prvlargew}). In the top panel, the sink is located at
  $\chizerohat=0.15$ ($\xzero=0.3$), i.e., close to the entry boundary,
  whereas the bottom panel is for $\chizerohat=0.45$ ($\xzero=0.9$),
  close to the reflective boundary. In both cases, the time-reversed
  probability is essentially the same, mostly uniform in time while
  exhibiting some gradient towards the reflective boundary. This
  situation corresponds to a saturated quantum diffusion regime and it is
  very different from the large-width limit displayed in
  \cref{fig:Prvlargew}.}
\label{fig:Prvsmallw}
\end{center}
\end{figure}

In \cref{fig:Prvsmallw}, we have plotted $G \sqrt{\rv{\Nzero}}
\Prv{\phi,\rv{N}\mid\phizero,\rv{\Nzero}}$ for $\chirhat=0.5$, which
corresponds to a well width equal to half the typical Brownian
excursion $G\sqrt{\rv{\Nzero}}$ at a given lifetime. The top and
bottom panels are for two different locations of the sink (the initial
condition of the forward process), $\xzero=0.3$ and $\xzero=0.9$,
respectively. In both cases, and when the field is not too close to
the entry boundary and to the sink, the distribution is uniform in
$\tau$ and exhibits a gradient from the entry to the reflective
boundary. A well width, measured with the values of $\chirhat$, which
is less than unity indeed implies that the diffusion domain is too
small for the natural Brownian excursion. As a result, at given
lifetime $\rv{\Nzero}$, the stochastic field has more than enough time
to explore the whole domain, and the extra time is actually spent in
bouncing against the reflective wall at $\phi=\phir$. An equivalent
reasoning is to remark that $\chirhat <1$ implies that $\rv{\Nzero} >
(\rv{\phir}/G)^2$, i.e., the lifetime of the processes is greater than
the typical diffusion time to explore the domain $\rv{\phir}$. We will
refer to this regime as a ``saturated quantum diffusion'': at any
time, the field can take any value within the well, with a probability
smoothly growing towards its maximum at $\phir$. Such a situation
cannot occur in an unbounded potential. A finite quantum well with
$\chirhat<1$ therefore exhibits stronger quantum effects than the
semi-infinite flat potential.

The limit $\chirhat \ll 1$ implies that $\qzero \to
0^{+}$. From the definition~\eqref{eq:deftheta2} of the second theta
function, in the limit $q\to0$, we have
\begin{equation}
\jtheta{2}{z,q} \simeq 2 q^{1/4}\cos(z), \qquad \djtheta{2}{z,q}
\simeq -2 q^{1/4} \sin(z),
\label{eq:thetaqnull}
\end{equation}
from which \cref{eq:Prvsolhat} gives
\begin{equation}
\lim_{\chirhat \ll 1} G \sqrt{\rv{\Nzero}} \,
\Prv{\phi,\rv{N}\mid\phizero,\rv{\Nzero}} = \dfrac{2}{\chirhat} \sin^2\qty(\dfrac{\pi}{2}\dfrac{\chihat}{\chirhat}).
\label{eq:Prvnull}
\end{equation}
This expression approaches very well the exact distribution plotted in
\cref{fig:Prvsmallw}, there is no dependence on $\tau$ and the
gradient towards the reflective boundary varies as $\sin^2(\pi
x/2)$. Let us further stress that \cref{eq:Prvnull} no longer has any
dependence on the sink location $\chizerohat$. The regime of saturated
quantum diffusion effectively erases any memory of the initial
conditions of the forward process.

\subsection{Dependence on the well width}

\begin{figure}
    \begin{subfigure}{\twofigw}
        \includegraphics*[width=\twofigw]{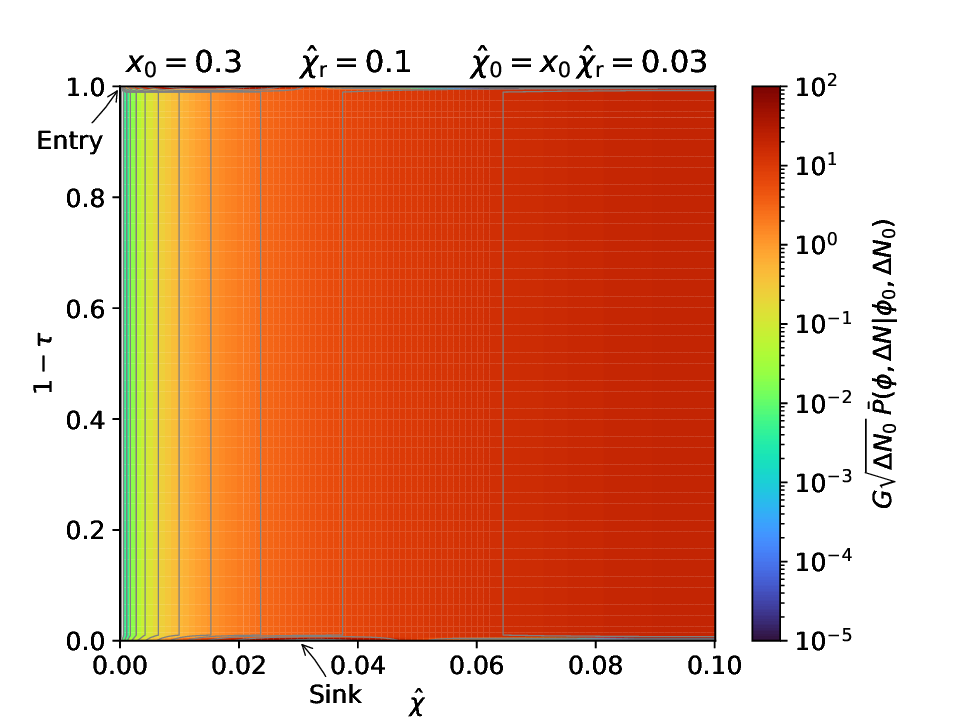}
        \includegraphics*[width=\twofigw]{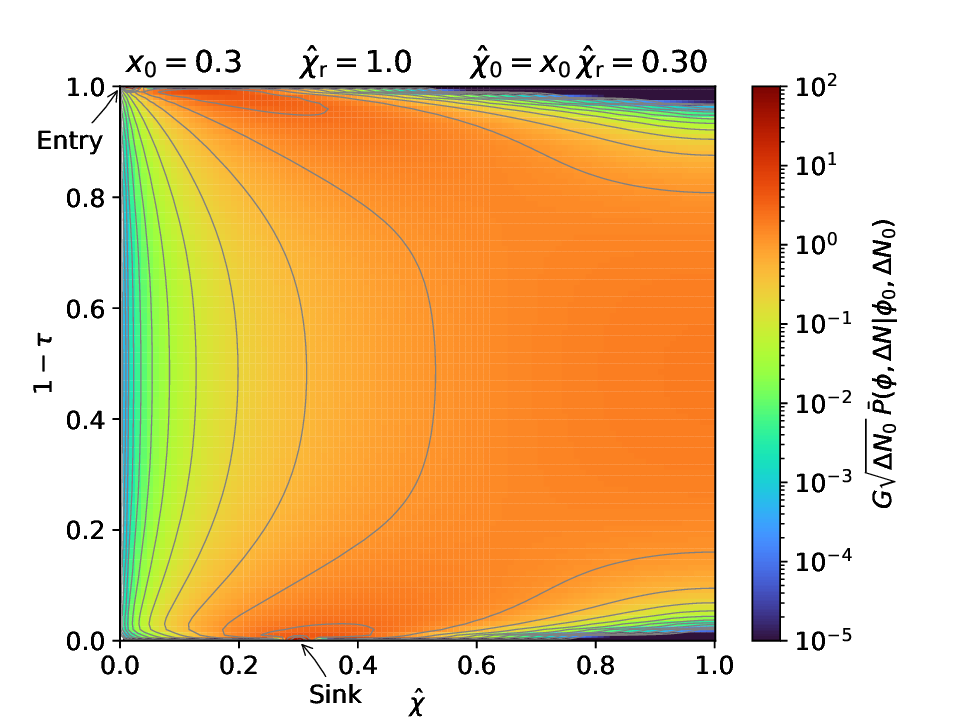}
        \includegraphics*[width=\twofigw]{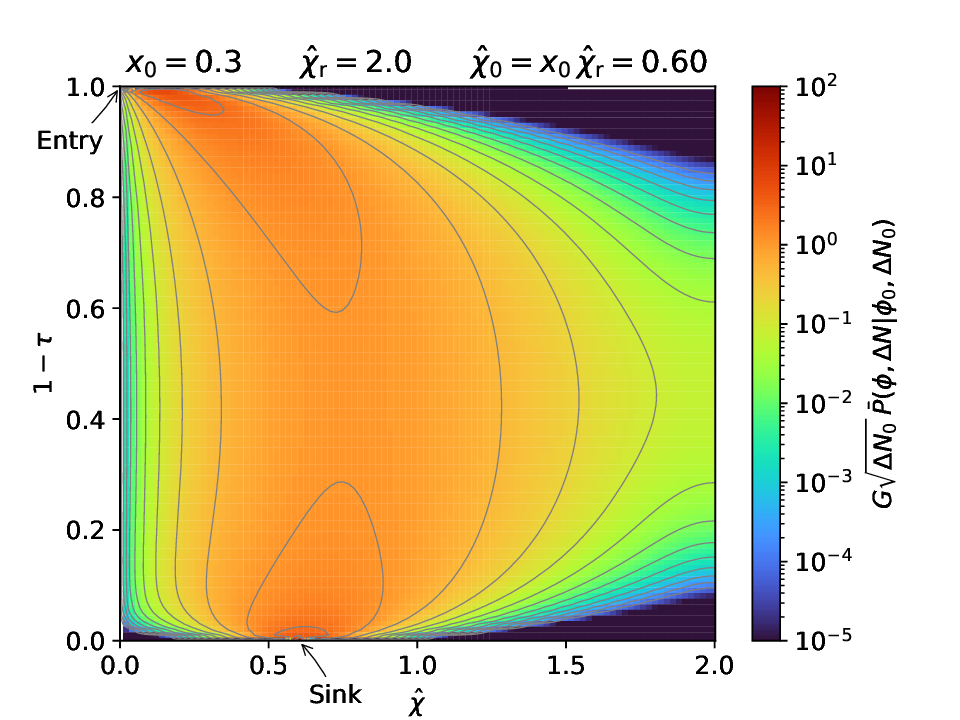}
        \caption{Flat well}
    \end{subfigure}
    \begin{subfigure}{\twofigw}
        \includegraphics*[width=\twofigw]{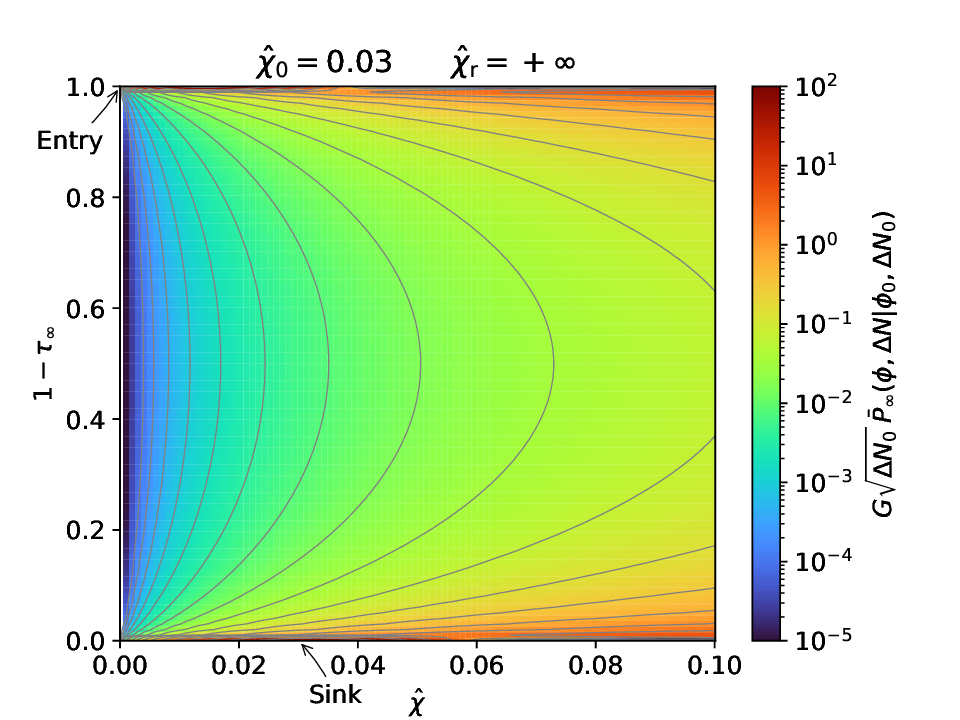}
        \includegraphics*[width=\twofigw]{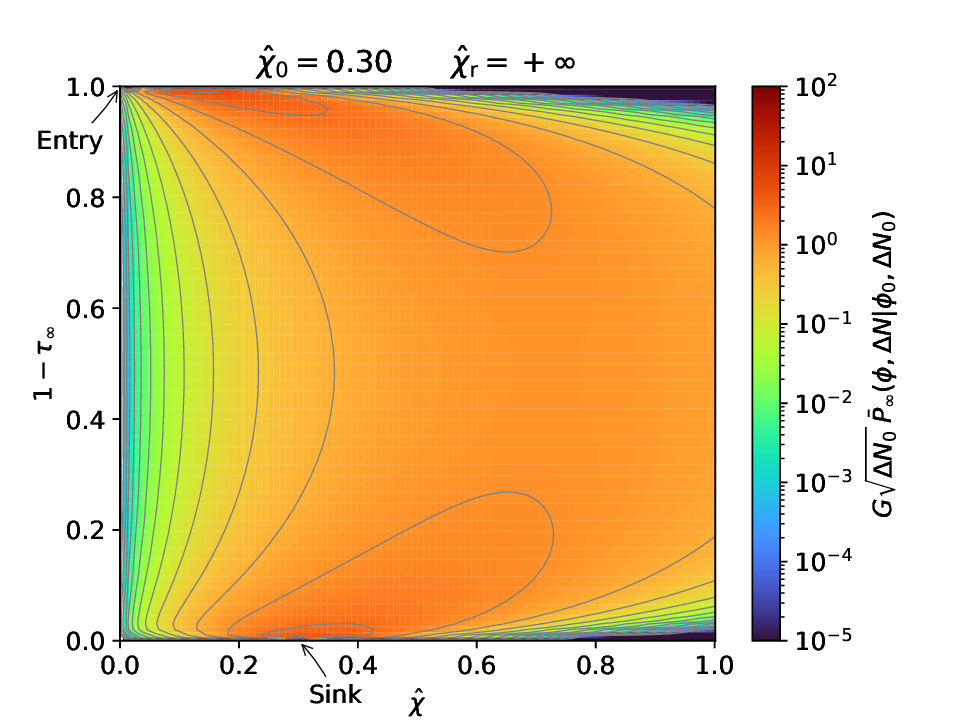}
        \includegraphics*[width=\twofigw]{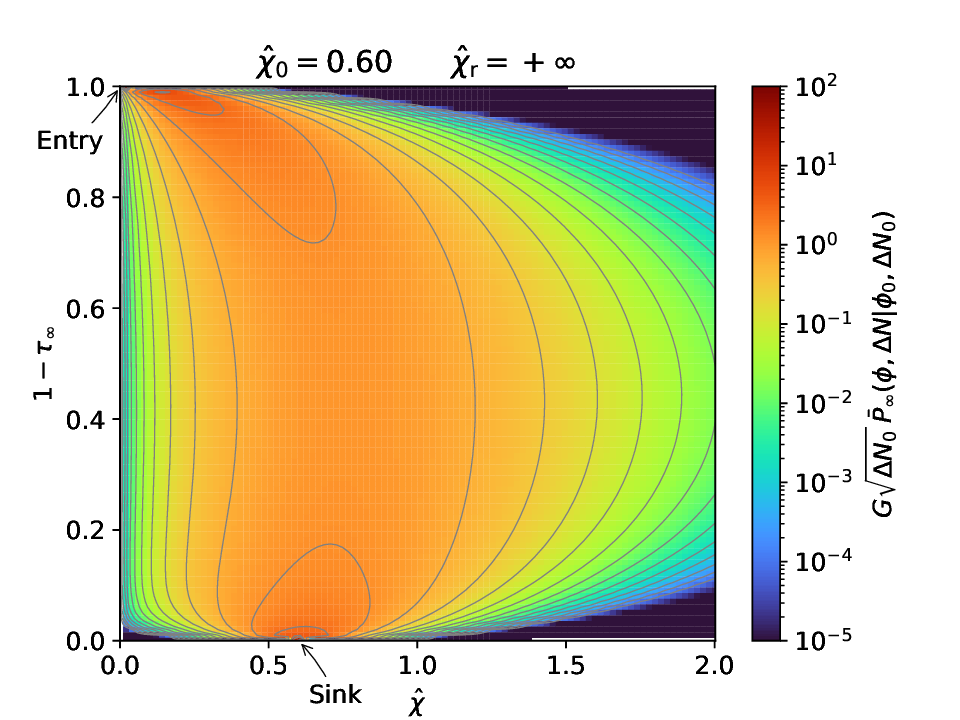}
        \caption{Semi-infinite flat potential}
    \end{subfigure}
    \caption{Contour plots of the rescaled time-reversed transition
  probability distribution, $G\sqrt{\rv{\Nzero}} \times
  \Prv{\phi,\rv{N}\mid\phizero,\rv{\Nzero}}$, for the flat well (left
  column) and for the semi-infinite flat potential (right column). The
  sink is located at $\chizerohat = \xzero \chirhat=0.3 \chirhat$,
  where $\chirhat$, the well width in units of Brownian excursion, is
  varied from $\chirhat=0.1$ to $\chirhat=2$ from top to bottom. The
  regime of saturated quantum diffusion at $\chirhat<1$ can only occur
  within the flat well. As the first row shows, the distribution is
  uniform in the flat well (top-left panel) and significantly differs from the one of
  the semi-infinite flat potential (top-right panel). By increasing
  $\chirhat$, quantum diffusion smoothly transits from ``saturated''
  to identical to the one of the semi-infinite flat potential.}
\label{fig:Prvallw}
\end{figure}

In \cref{fig:Prvallw}, we have plotted in the left column the rescaled
time-reversed probability distribution of \cref{eq:Prvsolhat} for
various values of $\chirhat$, the well width in units of Brownian
excursion, ranging from $\chirhat=0.1$ to $\chirhat=2.0$. The sink
location has been fixed to $\xzero=0.3$ in all these plots, so that
they reflect only changes in $\chirhat$. The right column of
\cref{fig:Prvallw} shows the rescaled time-reversed probability
distribution for the semi-infinite flat potential given by
\cref{eq:Prvinf}. The sink is located at the same position as in the
corresponding flat well, namely $\chizerohat=\xzero \chirhat = 0.3
\chirhat$, where $\chirhat$ refers to the width of the flat
well. Moreover, for the plots in the right column, the horizontal axis
has been artificially truncated at $\chihat=\chirhat$ -- the
corresponding maximal value in the flat well -- for ease of
comparison. The probability distribution in the semi-infinite
potential is well defined, and normalised, even for
$\chihat>\chirhat$, see Refs.~\cite{Blachier:2025tcq,Blachier:2025iwk}
for more details on the large field region. This figure illustrates
how, by increasing $\chirhat$ from values below to values above unity,
the system smoothly transits from the saturated quantum-diffusion
regime described by \cref{eq:Prvnull} to a diffusion regime
indistinguishable from that of the semi-infinite flat potential. For
$\chirhat \ll 1$, the diffusion does not depend on the value of
$\chizerohat$ and the saturated quantum diffusion generated by the
reflective boundary washes out the memory of the initial
conditions. In the opposite limit, $\chirhat\gg 1$, the reflective
boundary has no longer any effect and quantum diffusion is identical
to that occurring in the semi-infinite potential, it only depends on
the value of $\chizerohat$. Let us stress that this transition is most
obvious when field values are expressed in units of the typical
Brownian excursion, i.e., using $\chihat$, $\chirhat$ and
$\chizerohat$.

\section{Quantum-generated curvature perturbations}
\label{sec:curv}

In the previous section, we exactly solved time-reversed
stochastic inflation in the flat well. This probed the importance of
the parameter $\chirhat$, the well width in units of the typical
Brownian excursion $G\sqrt{\rv{\Nzero}}$, in determining the type of
quantum diffusion that occurs. In qualitative terms, we found
that for small $\chirhat$ the system exhibits a saturated regime of quantum diffusion, 
memory of the initial condition is washed out and the
reverse probability distribution is uniform in time and maximal on the
reflective boundary. In the large width limit, $\chirhat \gg 1$,
quantum diffusion is essentially the same as in the semi-infinite flat
potential, showing that the reflective boundary becomes irrelevant.

In this section, we use the reverse $\deltaN$ formalism described in
\cref{subsec:rev} to derive the probability distribution of the
generated curvature fluctuations, which first requires to determine
$\ev{\rv{N}}$.

\subsection{Mean number of reverse $e$-folds}

\begin{figure}
    \begin{subfigure}{\twofigw}
        \includegraphics[width=\twofigw]{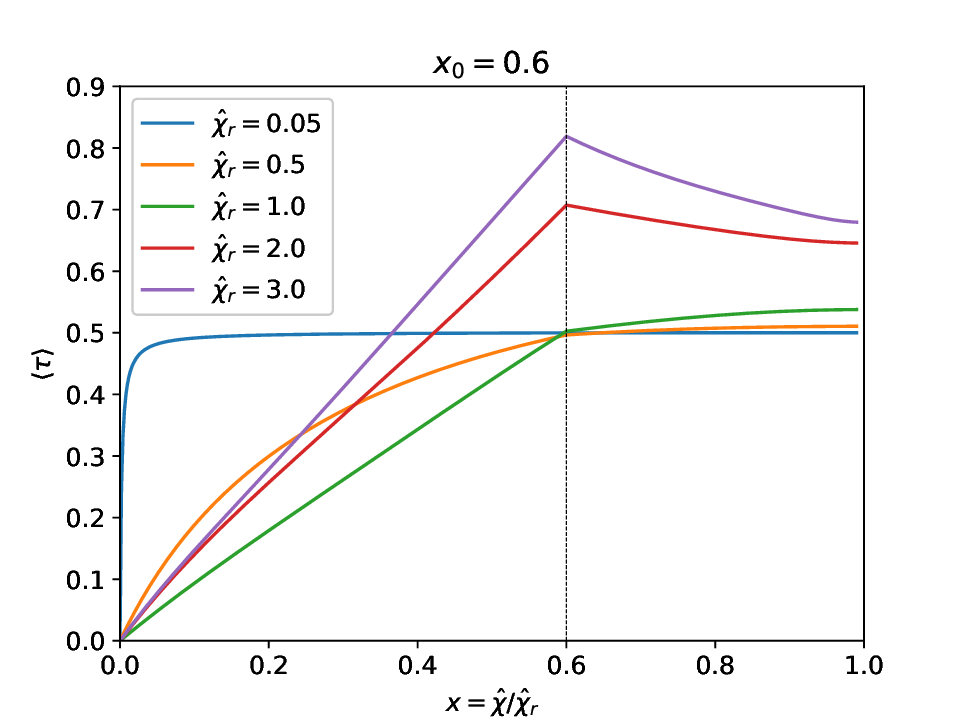}
        \includegraphics[width=\twofigw]{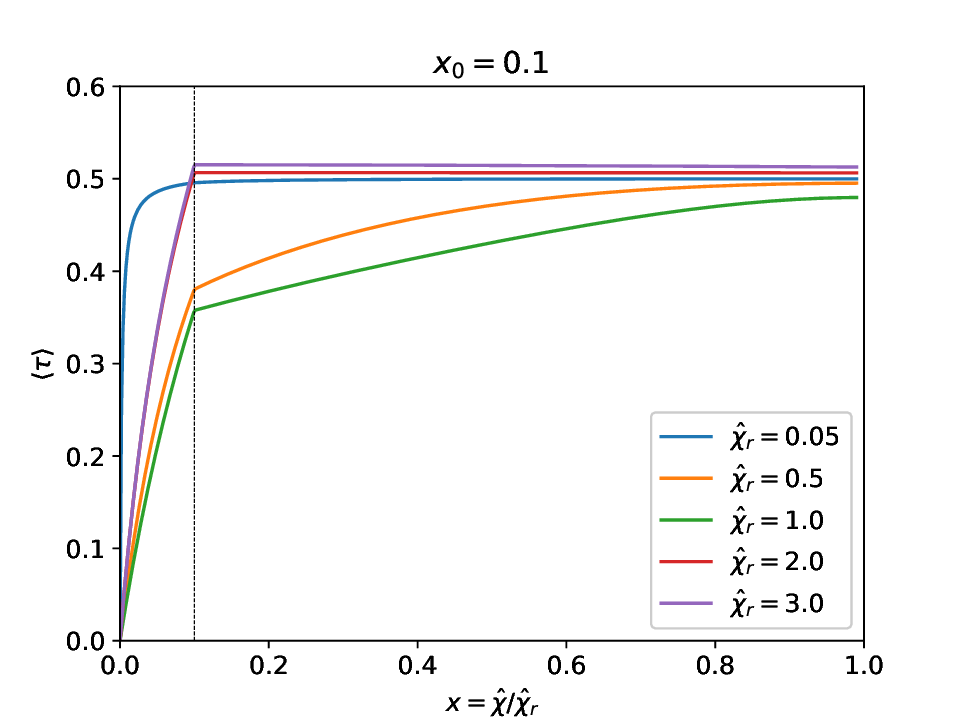}
        \caption{Flat well}
    \end{subfigure}
    \begin{subfigure}{\twofigw}
        \includegraphics[width=\twofigw]{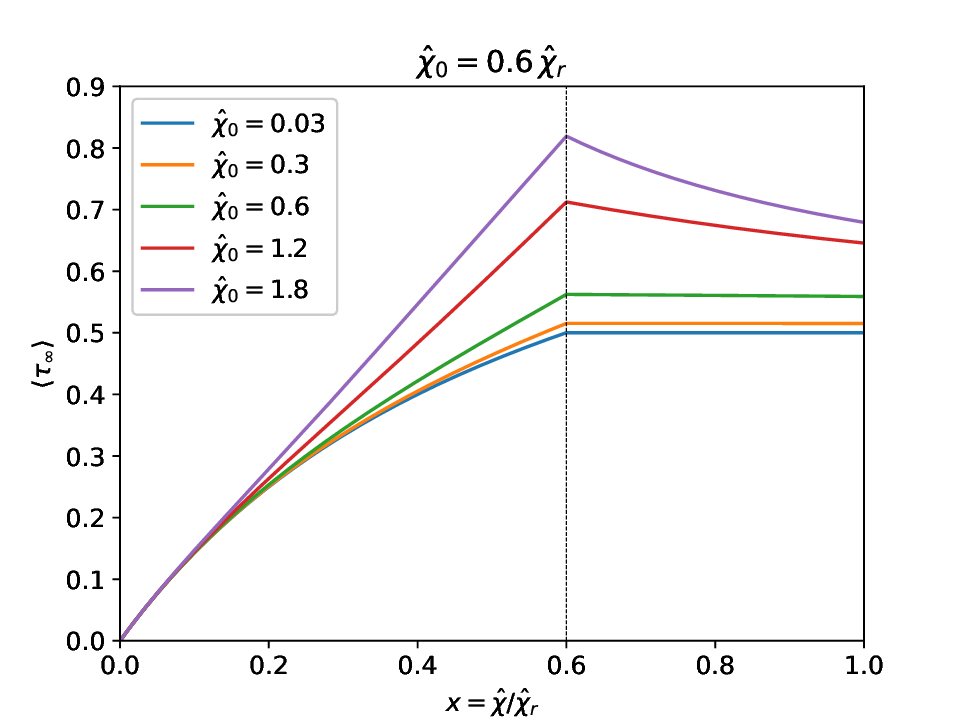}
        \includegraphics[width=\twofigw]{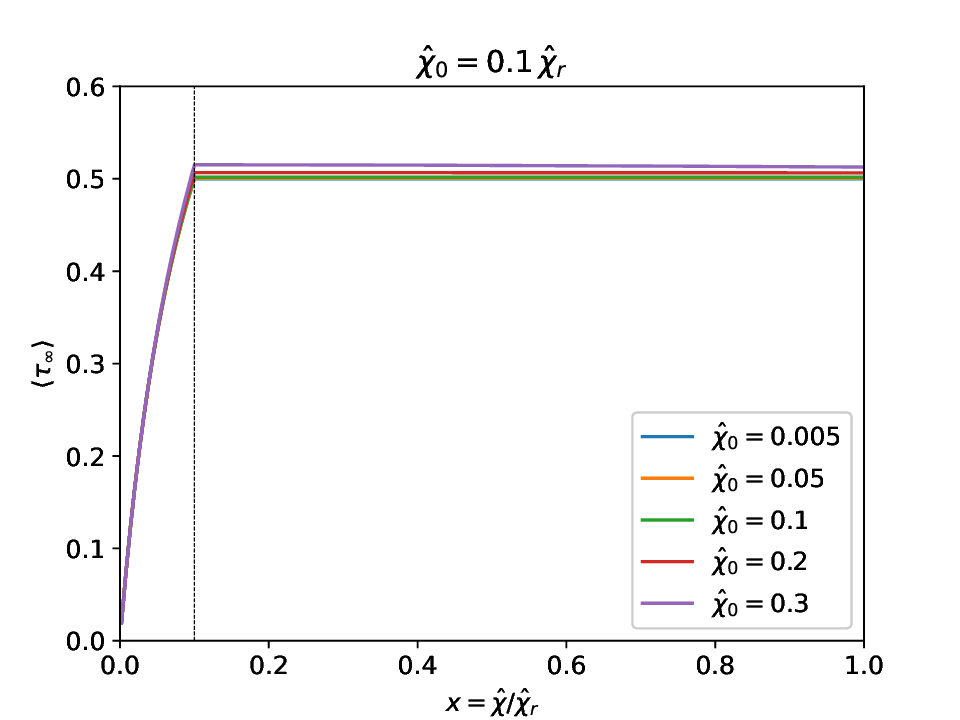}
        \caption{Semi-infinite flat potential}
    \end{subfigure}
    \caption{Mean value $\ev{\tau}$ of the number of reverse {\efolds}
      in units of the lifetime $\rv{\Nzero}$ as a function of field
      values $x=\chihat/\chirhat$. The left column corresponds to the
      flat well and should be compared to the right column, which
      corresponds to the semi-infinite flat potential (see
      \cref{eq:taumeaninf}). The top panels are for a sink located at
      $\xzero=0.6$ whereas the bottom ones are for
      $\xzero=0.1$. $\ev{\tau}$ in the flat well differs from the
      semi-infinite case only in the regime of saturated quantum
      diffusion $\chirhat \ll 1$, where one has $\ev{\tau} \simeq 1/2$
      in almost all the domain.}
\label{fig:taumean}
\end{figure}

The $\deltaN$ formalism allows one to map the curvature fluctuations
of the spacetime onto the fluctuations in the number of {\efolds}
through an adequate coordinate transformation. This very same
transformation allows us to describe stochastic inflation by the
manifestly homogeneous equation~\eqref{eq:langevin}. Therefore, a
given field value is associated with many possible realisations of the
spacetime and their associated number of {\efolds}. This can be
illustrated by considering slices of constant field values in
\cref{fig:Prvlargew,fig:Prvsmallw,fig:Prvallw}, each of which defines
a stochastic distribution for $\tau$, i.e., for $\rv{N}$. At fixed
lifetime $\rv{\Nzero}$, a given field value can indeed be reached by
many stochastic trajectories at different reverse {\efold} number
$\rv{N}$. In quantitative terms, one has
\begin{equation}
\Pof{\rv{N}\mid\phi,\phizero,\rv{\Nzero}} =
\dfrac{\Prv{\phi,\rv{N}\mid\phizero,\rv{\Nzero}}}{\displaystyle \int_0^{\rv{\Nzero}}
\Prv{\phi,\rv{N}\mid\phizero,\rv{\Nzero}} \dd{\rv{N}}}\,,
\label{eq:PofN}
\end{equation}
from which the mean number of {\efolds} is given by
\begin{equation}
\ev{\rv{N}} = \int_0^{\rv{\Nzero}} \rv{N}\,
\Pof{\rv{N}\mid\phi,\phizero,\rv{\Nzero}} \dd{\rv{N}}.
\label{eq:meanNdef}
\end{equation}
It is more convenient to work with the normalised {\efold} number
$\tau$ introduced in \cref{sec:solbyfp}. One has
\begin{equation}
  \ev{\rv{N}} = \ev{\tau} \rv{\Nzero},
\end{equation}
where, from \cref{eq:PofN,eq:meanNdef}, $\ev{\tau}$ is given by
\begin{equation}
\ev{\tau} = \dfrac{\displaystyle \int_0^1 \tau
  \Prv{\phi,\tau\rv{\Nzero}\mid\phizero,\rv{\Nzero}} \dd{\tau}}{\displaystyle \int_0^1
  \Prv{\phi,\tau\rv{\Nzero}\mid\phizero,\rv{\Nzero}} \dd{\tau}}\,.
\label{eq:taumeandef}
\end{equation}
Any common factor between the numerator and the denominator that do
not depend on $\tau$ cancel out, and from the time-reversed solution
of \cref{eq:Prvsolhat}, one has
\begin{equation}
\eval{\ev{\tau}}_{x,\xzero,\chirhat} = \dfrac{\displaystyle \int_0^1
  \tau \, \djtheta{2}{\dfrac{\pi}{2}
    x,\qzero^\tau}\qty[\jtheta{2}{\pi
      \dfrac{\xzero-x}{2},\qzero^{1-\tau}} - \jtheta{2}{\pi
      \dfrac{\xzero+x}{2},\qzero^{1-\tau}}] \dd{\tau} }{\displaystyle \int_0^1\djtheta{2}{\dfrac{\pi}{2}
    x,\qzero^\tau}\qty[\jtheta{2}{\pi
      \dfrac{\xzero-x}{2},\qzero^{1-\tau}} - \jtheta{2}{\pi
      \dfrac{\xzero+x}{2},\qzero^{1-\tau}}] \dd{\tau} }\,.
\label{eq:taumean}
\end{equation}

These integrals cannot be straightforwardly performed and there is no
closed-form expression for $\ev{\tau}$ in general. However, since we
showed in \cref{sec:solbyfp} that, in the large width limit,
$\Prv{\phi,\rv{N}\mid\phizero,\rv{\Nzero}}$ becomes identical to its
counterpart in the semi-infinite potential, \cref{eq:taumeandef}
ensures that $\ev{\tau}$ will also be given by its semi-infinite
analogue. Therefore, without performing any calculation, from
Ref.~\cite{Blachier:2025tcq} we have
\begin{equation}
\lim_{\chirhat \gg 1} \ev{\tau} = \sqrt{\dfrac{\pi}{2}} \, \chihat \,
\exp \qty( \frac{\chizerohat^2}{2} ) \, \dfrac{\erf\left(\dfrac{2 \chihat +
    \chizerohat}{\sqrt{2}}\right) - \erf\left(\dfrac{\chihat +
    \left|\chihat-\chizerohat\right|}{\sqrt{2}}\right)}{e^{-\chihat\left(\left|\chihat-\chizerohat\right|
    + \chihat-\chizerohat\right)} -
  e^{-2\chihat\left(\chihat+\chizerohat\right)}} \equiv \ev{\tau_\infty}.
\label{eq:taumeaninf}
\end{equation}
For $\chirhat \ll 1$, in the saturated quantum-diffusion regime, the
approximated distribution derived in \cref{eq:Prvnull} no longer
depends on $\tau$. One therefore immediately gets
\begin{equation}
\lim_{\chirhat \ll 1}\ev{\tau} = \dfrac{1}{2}\,,
\end{equation}
which is confirmed by a visual inspection of \cref{fig:Prvsmallw}.

For intermediate values of $\chirhat$, one has to rely on a numerical
integration of \cref{eq:taumean}. For this purpose, we have used the
numerical integrators provided by the {\SUNDIALS}
library~\cite{gardner2022sundials, hindmarsh2005sundials}. Fast and
accurate evaluations of the Jacobi theta functions have been provided
by the {\FLINT} project\footnote{\url{https://flintlib.org}}, using
ball arithmetic~\cite{johansson2017flint}. In \cref{fig:taumean}, we
have plotted the dependence of $\ev{\tau}$ on $x=\chihat/\chirhat$,
the field value in units of the well width, for various values of the
well width $\chirhat$. The left column shows the exact result,
computed using \cref{eq:taumean}, for two sink positions $\xzero=0.6$
(top) and $\xzero=0.1$ (bottom). The right column shows the same
quantity in the semi-infinite flat potential, given by
\cref{eq:taumeaninf}. Already for $\chirhat \gtrsim 1$, there is
almost no difference between the exact result of $\ev{\tau}$ and the
semi-infinite one. Most of the differences arise when the reflective
boundary interferes with the stochastic diffusion, i.e., for $\chirhat
< 1$. As discussed earlier, quantum diffusion is efficient in the flat
well and this pushes $\ev{\tau}$ much closer to $1/2$ than it would be
in the absence of a reflecting boundary. This is well illustrated by
comparing the left and right panels of \cref{fig:taumean} for the
curves labelled $\chirhat=0.05$. However, at fixed $\xzero$,
considering small values of $\chirhat$ also implies small values of
$\chizerohat = \xzero\chirhat$. As shown in
Refs.~\cite{Blachier:2025tcq,Blachier:2025iwk}, the small
$\chizerohat$ limit in the semi-infinite flat potential also induces
strong quantum diffusion but only for $\chihat > \chizerohat$. As
such, for the semi-infinite flat potential, one also has
$\ev{\tau}\simeq 1/2$ for $\chihat>\chizerohat$, which is actually
most of the domain for small values of $\chizerohat$. This behaviour
justifies why, in the end, most of the differences between the left
and right column of \cref{fig:taumean} are essentially visible only at
$x<\xzero$.

\subsection{Curvature perturbation at given lifetime}
\label{sec:zetahat}

\begin{figure}
    \begin{subfigure}{\twofigw}
         \includegraphics[width=\twofigw,clip=false]{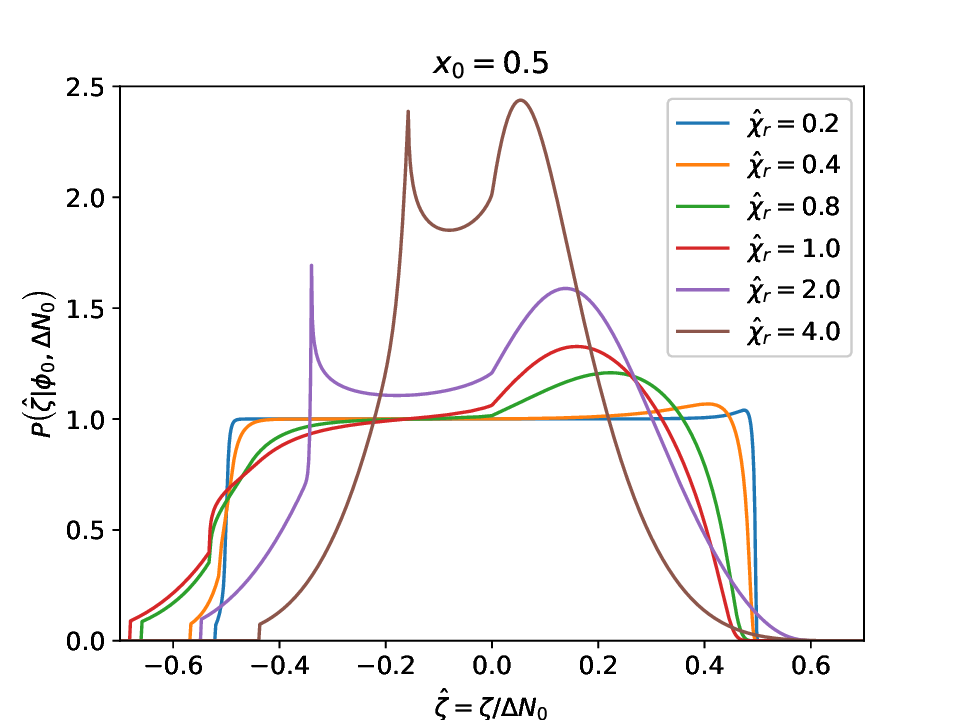}
         \caption{Flat well}
    \end{subfigure}
    \begin{subfigure}{\twofigw}
        \includegraphics[width=\twofigw,clip=false]{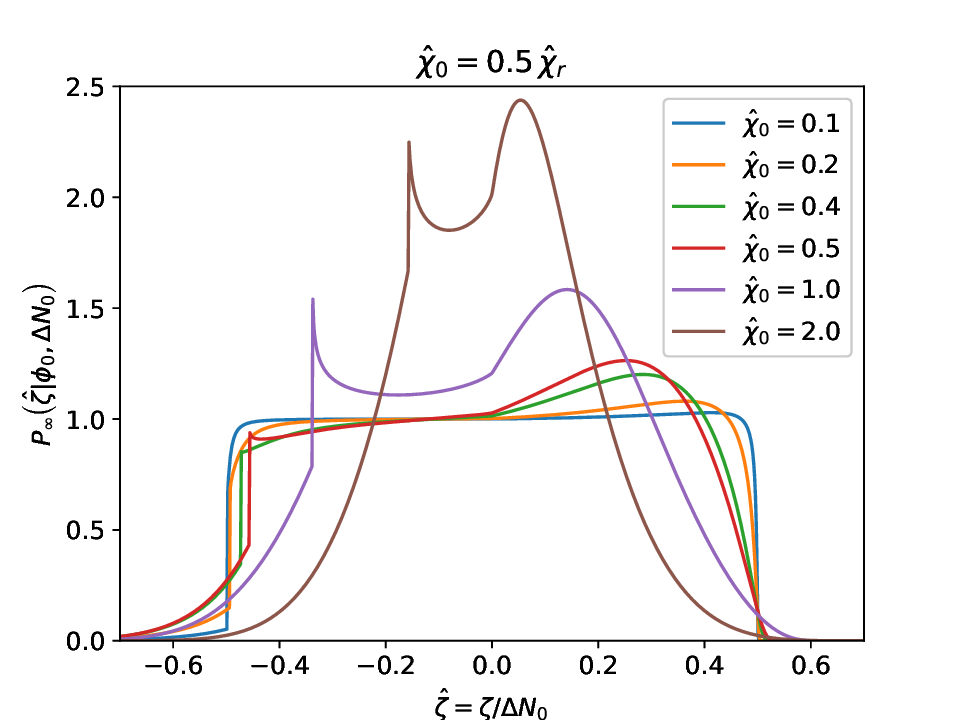}
        \caption{Semi-infinite flat potential}
    \end{subfigure}
\caption{Probability distribution of $\zetahat=\zeta/\rv{\Nzero}$, the
  curvature fluctuation in units of the lifetime, for the flat well
  (left) and for the semi-infinite flat potential (right). The sink is
  located in the middle of the well at $\xzero=1/2$ and the different
  curves show the effect of changing $\chirhat$, the well width in
  units of the typical Brownian excursion. For $\chirhat \ll 1$ the
  distribution approaches the rectangle function while for $ \chirhat
  \gg 1$ it is indistinguishable from the semi-infinite one.}
\label{fig:pzetahatcomp}
\end{figure}

As can be seen from \cref{eq:Prvsolhat}, after performing the replacement
$\rv{N} \to \ev{\rv{N}} - \zeta$ in \cref{eq:Pofzetaphidef}, all
dependence on the lifetimes $\rv{\Nzero}$ enters through the quantity
$\tau$ and its mean value $\ev{\tau}$ defined in \cref{eq:taumean}. It
is therefore more convenient to introduce the new quantity
\begin{equation}
\zetahat \equiv \dfrac{\zeta}{\rv{\Nzero}}\,,
\label{eq:zetahatdef}
\end{equation}
which is the curvature fluctuation measured in units of the
lifetime. From \cref{eq:Pofzetaphidef}, the joint probability
distribution of $\zetahat$ and $\phi$ is, in fact, given by the
time-reversed distribution
\begin{equation}
\Pof{\phi,\zetahat\mid\phizero,\rv{\Nzero}} = \Prv{\phi,\tau=\ev{\tau}-\zetahat\mid\phizero,\rv{\Nzero}}.
\label{eq:P:zetahat:prop:Pbar}
\end{equation}
As we show in Appendix~\ref{sec:normalisation}, due to the mass conservation property of the
Fokker-Planck equation, $\Prv{\phi,\tau=\ev{\tau}-\zetahat\mid\phizero,\rv{\Nzero}}$ is
indeed normalised with respect to both $\phi$ and $\zetahat$. From
\cref{eq:Prvsolhat}, the joint probability distribution of $\phi$
and $\zetahat$, at given lifetime, is therefore given by
\begin{equation}
  \begin{aligned}
\Pof{\phi,\zetahat\mid\phizero,\rv{\Nzero}} & = \dfrac{1}{G \sqrt{\rv{\Nzero}}}
\dfrac{1}{2 \chirhat} \dfrac{\djtheta{2}{\dfrac{\pi}{2}\dfrac{\chihat}{\chirhat}
    ,\qzero^{\ev{\tau} -
      \zetahat}}}{\djtheta{2}{\dfrac{\pi}{2}\dfrac{\chizerohat}{\chirhat},\qzero}}
\qty[\heaviside{\ev{\tau} - \zetahat} - \heaviside{\ev{\tau} -
    \zetahat - 1}]
\\ & \times \qty[
  \jtheta{2}{\dfrac{\pi}{2}\dfrac{\chizerohat-\chihat}{\chirhat},\qzero^{1-\ev{\tau}+\zetahat}}
  - \jtheta{2}{\dfrac{\pi}{2}\dfrac{\chizerohat+\chihat}{\chirhat},\qzero^{1-\ev{\tau}+\zetahat}} ],
  \end{aligned}
  \label{eq:Pphizetahat}
\end{equation}
where $\ev{\tau}$ is given by \cref{eq:taumean}. The Heaviside
functions appearing in the first line of \cref{eq:Pphizetahat}
enforce the condition $0<\rv{N} < \rv{\Nzero}$, which gets translated
into $0< \ev{\tau} - \zetahat < 1$. They act as window functions on
the field values, selecting only the domains compatible with a given
$\zetahat$. Let us stress again that, up to an overall
$G\sqrt{\rv{\Nzero}}$ factor on the joint probability, using
$\zetahat$ instead of $\zeta$ allows us to remove any explicit
dependence on the lifetime $\rv{\Nzero}$. 

In order to obtain the distribution of $\zetahat$, one has to
marginalise over the field values $\phi$. From the change of variable
$\phi \to x$, using \cref{eq:chihatdef,eq:xotherdef}, the factors $G
\sqrt{\rv{\Nzero}}$ and $\chirhat$ disappear and one obtains
\begin{equation}
  \begin{aligned}
\Pof{\zetahat\mid\phizero,\rv{\Nzero}} &= \dfrac{1}{2\djtheta{2}{\dfrac{\pi}{2}
    \xzero,\qzero}} \int_0^1 \dd{x} \qty[\heaviside{\ev{\tau} - \zetahat} - \heaviside{\ev{\tau} -
    \zetahat - 1}] \\ & \times
\djtheta{2}{\dfrac{\pi}{2}x,\qzero^{\ev{\tau}-\zetahat}} 
\Bqty{
  \jthetab{2}{\dfrac{\pi}{2}\qty(\xzero-x),\qzero^{1-\ev{\tau}+\zetahat}}
  -
  \jthetab{2}{\dfrac{\pi}{2}\qty(\xzero+x),\qzero^{1-\ev{\tau}+\zetahat}} },
\end{aligned}
\label{eq:pzetahat}
\end{equation}
which is already normalised to unity for $\zetahat$ as mentioned earlier. The only
remaining dependence of this expression on $\chirhat$, the well width
in units of the typical Brownian excursion at given lifetime, is within
the function $\qzero=\exp[-\pi^2/(2\chirhat^2)]$. As explained
earlier, the semi-infinite flat-potential limit has to be recovered
for $\chirhat \gg 1$, i.e., for $\qzero \to 1^{-}$. This is indeed the
case as shown in \cref{fig:pzetahatcomp}. For a sink located in the
middle of the well, $\xzero=1/2$, we have plotted on the left the
exact distribution for $\zetahat$, as obtained by a numerical
integration of \cref{eq:pzetahat}, for various values of
$\chirhat$. The right panel of this figure shows
$\Pname_\infty(\zetahat\mid\phizero,\Nzero)$, which is the
distribution of $\zetahat$ obtained in Ref.~\cite{Blachier:2025tcq,
  Blachier:2025iwk} for the semi-infinite flat potential (also
determined numerically). As before, changing $\chirhat$ at fixed
$\xzero$ in the flat well also modifies the location of the sink
$\chizerohat = \xzero \chirhat$, and the corresponding values are
reported in the legend of \cref{fig:pzetahatcomp}. We recover the fact
that, as soon as $\chirhat \gtrsim 1$, there is no visible difference
between the flat well and the semi-infinite flat potential. However,
this figure also shows that, even for $\chirhat < 1$, the two
distributions remain very similar as both converge towards the
rectangle function. This is due to the similarity of the long-lifetime
limits in these two setups. The first concerns the semi-infinite
potential in which taking the sink at $\chizerohat\ll 1$ produces the
so-called diffusion regime, $\rv{\Nzero} \gg (\rv{\phizero}/G)^2$, in
which $\ev{\tau}\simeq1/2$ and
$\Pname_\infty(\zetahat\mid\phizero,\Nzero)$ converges towards the
rectangle distribution~\cite{Blachier:2025tcq, Blachier:2025iwk}. The
second concerns the flat well, for which taking the small width limit,
$\chirhat \ll 1$, generates the saturated quantum-diffusion regime
discussed in \cref{sec:saturated} where one has $\rv{\Nzero} \gg
(\rv{\phir}/G)^2$ and $\ev{\tau}\simeq 1/2$. Moreover, using the
approximated time-reversed distribution derived in \cref{eq:Prvnull}
for $\chirhat \ll 1$, the integral appearing in \cref{eq:pzetahat} is
trivial and one gets
\begin{equation}
\lim_{\chirhat \ll 1} \Pof{\zetahat\mid\phizero,\rv{\Nzero}} =
\heaviside{\dfrac{1}{2} - \zetahat} - \heaviside{-\dfrac{1}{2} -
  \zetahat} \equiv \rect\qty(\zetahat).
\label{eq:rectlimit}
\end{equation}
Then, as far as $\zetahat$ is concerned, the flat well differs from
the semi-infinite potential only for widths of the same order as the
typical Brownian excursion, i.e., $\chirhat \simeq 1$.

\begin{figure}
\begin{center}
  \includegraphics*[width=\onefigw,clip=false]{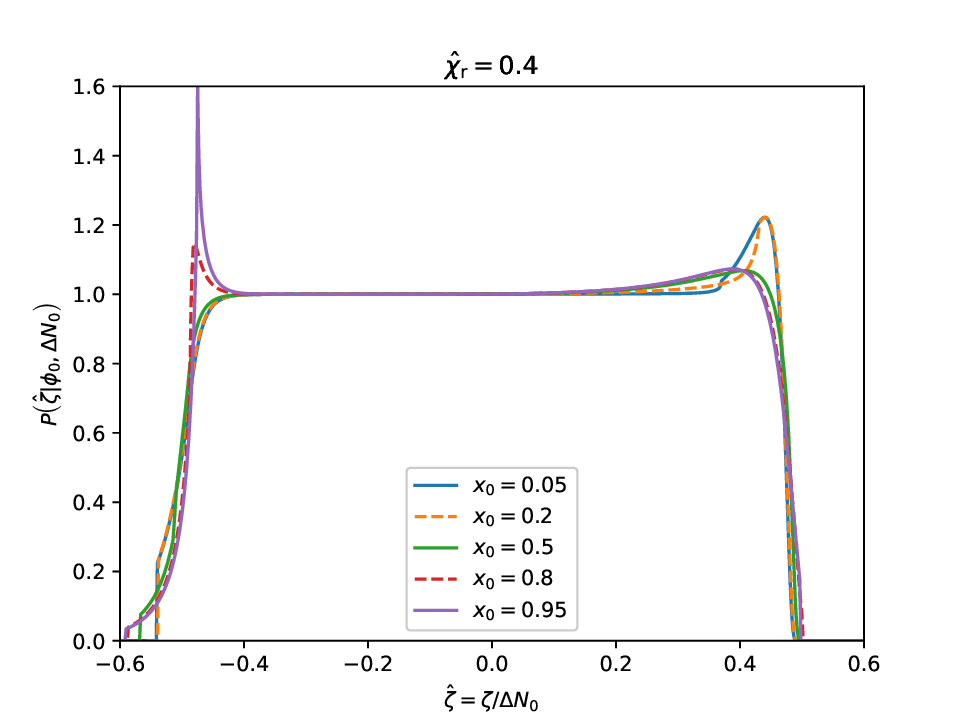}
  \caption{Probability distribution of $\zetahat$, the curvature
    fluctuation in units of the lifetime, in a flat well having a
    width $\chirhat=0.4$. The different curves show the effect of
    changing the sink location $\xzero$ (the initial condition of the
    forward process). For $\xzero$ close to the exit or to the
    reflective boundary, there is an enhancement of the probability
    around $\zetahat \simeq \pm 1/2$. Otherwise, the distribution
    remains mostly unaffected.}
  \label{fig:pzetahadxo}
\end{center}
\end{figure}

In \cref{fig:pzetahadxo}, we have plotted the probability distribution
of $\zetahat$ for a width fixed to $\chirhat=0.4$ and for different
values of the sink location $\xzero$ (the initial condition of the
forward process). The value of $\chirhat$ has been chosen to yield a
distribution intermediate between the rectangle and the semi-infinite
limits. As can be seen in this figure, the effects of $\xzero$ are
significant only when it lies very close to the sink or to the entry
boundary. In these cases, the probability of finding $\zetahat$ is
enhanced around $\zetahat \simeq \pm 1/2$. For other values of
$\xzero$, \cref{fig:pzetahadxo} shows that the distribution remains
unaffected so that the plots of \cref{fig:pzetahatcomp} are actually
representative of the generic case.

\subsection{Probability distribution of the curvature fluctuations}

In order to derive the final probability distribution of $\zeta$, one
still has to marginalise the distribution of $\zetahat$ derived in the
previous section over all possible lifetimes $\rv{\Nzero}$. Paying
attention to the definition of $\zetahat$ in \cref{eq:zetahatdef}, one
has, by virtue of probability conservation,
\begin{equation}
\Pof{\zeta\mid\phizero,\rv{\Nzero}} = \dfrac{1}{\rv{\Nzero}} \Pof{\zetahat\mid\phizero,\rv{\Nzero}}  ,
\label{eq:proba_conserv}
\end{equation}
and \cref{eq:Pofzetadef} reads
\begin{equation}
\Pof{\zeta\mid\phizero} = \int_0^{+\infty}
\Pof{\zetahat\mid\phizero,\rv{\Nzero}} \dfrac{\Plt{\rv{\Nzero}\mid\phizero}}{\rv{\Nzero}}
\dd{\rv{\Nzero}}.
\label{eq:pzeta}
\end{equation}
As shown in the previous section, at given $\xzero$,
$\Pname(\zetahat\mid\phizero,\rv{\Nzero})$ depends on $\rv{\Nzero}$ only
through $\zetahat$ and $\chirhat$. Let us change the integration
variable from $\rv{\Nzero}$ to $\chirhat = \chir/\sqrt{\rv{\Nzero}}$
at fixed $\chir = \rv{\phir}/G$, the well width in units of the
diffusion coefficient. Making explicit all dependencies on the
integration variable, one gets, using \cref{eq:Plt,eq:fwPfpt},
\begin{equation}
\Pof{\zeta\mid\phizero} = -\dfrac{\pi}{4\chir^2} \int_0^{+\infty}
\dfrac{\Pof{\left. \zetahat = \dfrac{\zeta}{\chir^2}\times
    \chirhat^2 \, \right \vert \phizero,\hcancel{\rv{\Nzero}}}}{\chirhat}
2\djtheta{2}{\dfrac{\pi}{2}\xzero,e^{-\frac{\pi^2}{2\chirhat^2}}} \dd{\chirhat}.
\end{equation}

The crossed out $\hcancel{\rv{\Nzero}}$ is used to recap that
$\Pname(\zetahat|\phizero,\rv{\Nzero})$ does not explicitly depend on
the lifetime, as can be seen in \cref{eq:pzetahat}. This expression
proves that the probability distribution of $\zeta$, multiplied by
$\chir^2$, is only a functional of $\zeta/\chir^2$ and $\xzero$. From
the exact expression of $\Pof{\zetahat\mid\phizero,\rv{\Nzero}}$ in
\cref{eq:pzetahat}, we see that the normalisation factor involving $2
\djtheta{2}{\pi \xzero/2,\qzero}$ cancels out, therefore leading to
\begin{equation}
  \begin{aligned}
& \chir^2 \Pof{\zeta\mid\phizero}  = \dfrac{\pi}{4}
\int_0^{\frac{\chir}{\sqrt{|\zeta|}}} \dfrac{\dd{\chirhat}}{\chirhat}
\int_0^1 \dd{x}  \qty[\heaviside{\ev{\tau} - \dfrac{\zeta}{\chir^2} \chirhat^2} - \heaviside{\ev{\tau} -
    \dfrac{\zeta}{\chir^2} \chirhat^2 - 1}] \\ & \times
\djtheta{2}{\dfrac{\pi}{2}x,\qzero^{\ev{\tau}-\frac{\zeta}{\chir^2} \chirhat^2}}
\Bqty{
  \jthetab{2}{\dfrac{\pi}{2}\qty(\xzero+x),\qzero^{1-\ev{\tau}+\frac{\zeta}{\chir^2} \chirhat^2}} 
  - \jthetab{2}{\dfrac{\pi}{2}\qty(\xzero-x),\qzero^{1-\ev{\tau}+\frac{\zeta}{\chir^2} \chirhat^2}}
  },
  \end{aligned}
  \label{eq:pzetaexplicit}
\end{equation}
where we recap that $\ev{\tau}$ is a function of $(x,\xzero,\chirhat)$
given by the integrals in \cref{eq:taumean}, and $\qzero =
\exp[-\pi^2/(2\chirhat^2)]$. This expression involves three levels of
nested integrals, the innermost ones entering the expression of
$\ev{\tau}$, making its numerical evaluation challenging. As already
mentioned, $\ev{\tau}$ has been evaluated with direct and fast
integration methods based on the {\SUNDIALS} library. For the
two-dimensional integral over $(\chirhat,x)$ appearing in
\cref{eq:pzetaexplicit}, we have written a dedicated modern {\Fortran}
code using the Monte-Carlo integrator
{\CUBA}~\cite{Hahn:2004fe,Hahn:2014fua} and parallelised using the
message passing interface (\texttt{MPI}) complemented with
\texttt{OpenMP} directives.

\begin{figure}
\begin{center}
  \includegraphics*[width=\onefigw]{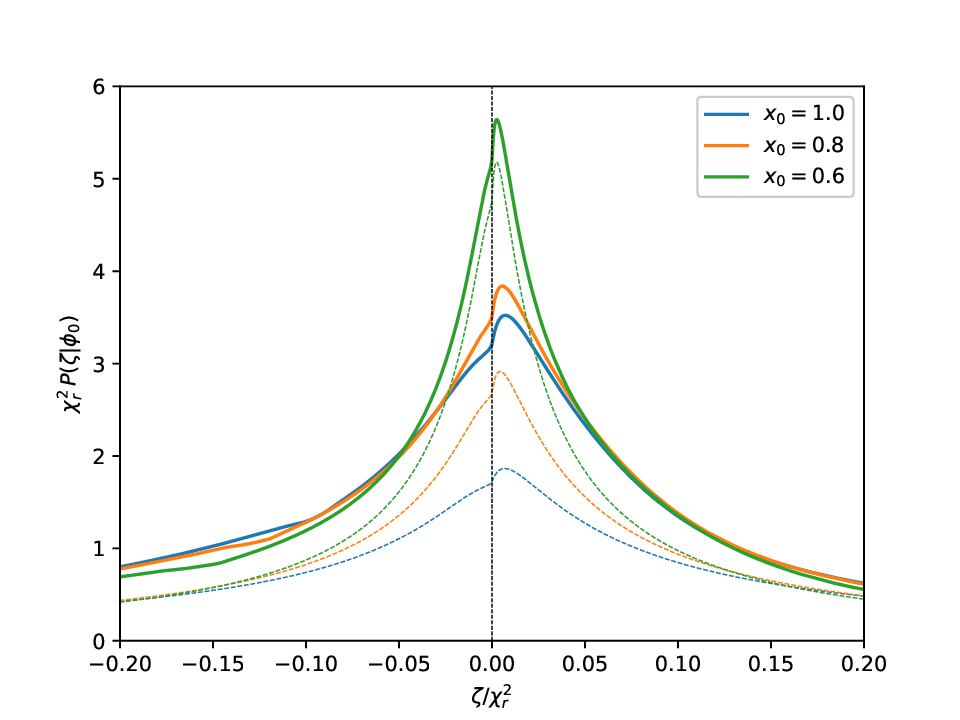}
  \includegraphics*[width=\onefigw]{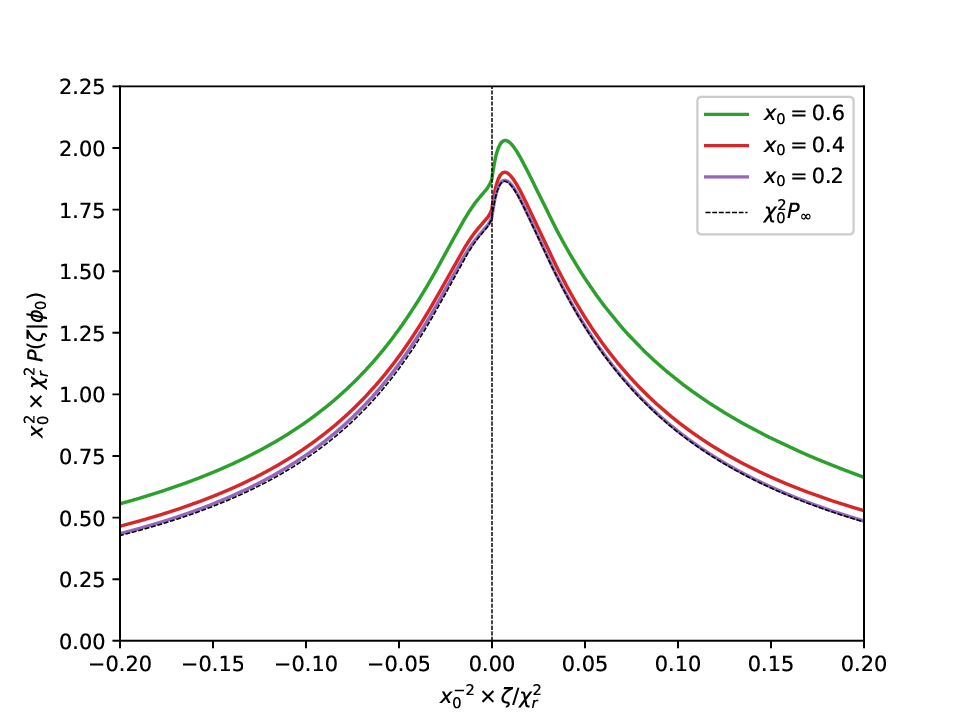}
\caption{Normalised probability distribution for the curvature
  fluctuations $\zeta$ in the quantum well (solid lines), for various
  values of the sink location $\xzero$. The upper panel shows $\chir^2
  \times \Pof{\zeta\mid\phizero}$ as a function of $\zeta/\chir^2$, which
  makes explicit any dependence on $\chir=\rv{\phir}/G$. The dashed
  curves represent the corresponding probability distribution in
  the semi-infinite flat potential. For $\xzero$ close to unity, they
  are different in amplitude while having the same overall shape in
  the domain $\abs{\zeta} < \chir^2$. For $\xzero$ small, both
  distributions become indistinguishable, as expected from
  \cref{eq:magicmath}. The lower panel shows $\xzero^2 \times
  \chir^2\Pof{\zeta\mid\phizero}$ as a function of $\xzero^{-2}
  \times \zeta/\chir^2$ to show how the flat well distributions (solid
  curves) converge towards the semi-infinite one (dashed curve). See
  \cref{fig:pzetatails} for the tails of the distribution.}
\label{fig:pzeta}
\end{center}
\end{figure}

The resulting normalised probability distribution for $\zeta$ has been
plotted in \cref{fig:pzeta}, for different values of the sink location
$\xzero$ and in the small-fluctuation domain. Notice that the
dependence on $\chir=\rv{\phir}/G$, the well width in units of the
diffusion coefficient, is explicit, so that we have represented in the
upper panel $\chir^2 \Pof{\zeta\mid\phizero}$ (solid lines) as a
function of $\zeta/\chir^2$. The dashed thin lines in the same figure,
which are quite close to the exact result (solid lines) for $\xzero <
0.6$, show the corresponding probability distribution derived in
Refs.~\cite{Blachier:2025tcq, Blachier:2025iwk} for the semi-infinite
flat potential. Indeed, in these works, it was shown that $\chizero^2
\Pinf{\zeta\mid\phizero}$ is a functional of $\zeta/\chizero^2$
only. Since we have shown here that, for $\chirhat \gg 1$, the
reflective boundary has no effect on quantum diffusion, the
probability distribution of $\zeta$ in the flat well should match that
derived in the semi-infinite flat potential in this limit. When this
occurs, the same reasoning also gives the functional form of
$\Pof{\zeta\mid\phizero}$, so that one must have
\begin{equation}
\chir^2 \Pof{\zeta\mid\phizero} \simeq \dfrac{1}{\xzero^2} \chizero^2
\Pinf{\left.\dfrac{\zeta}{\xzero^2} \right| \phizero},
\label{eq:magicmath}
\end{equation}
where use of \cref{eq:xotherdef} has been made. When do we expect this
relation to be valid? From \cref{eq:pzetaexplicit}, we see that the
domain $\chirhat \gg 1$ is probed by the integral provided the upper
bound is large enough, and a necessary condition is to have
$\abs{\zeta}/\chir^2 \ll 1$. In other words, for any well width, we
expect the curvature fluctuations in the flat well to be similar to
the ones in the semi-infinite flat potential when they are small
enough. Conversely, we therefore expect the tails of the probability
distribution to be very different. This is confirmed in
\cref{fig:pzeta} where, in the bottom panel, we have plotted the
distribution $\xzero^2 \times \chir^2 \Pof{\zeta\mid\phizero}$ as a
function of $\xzero^{-2} \times \zeta/\chir^2$. If the
relation~\eqref{eq:magicmath} holds, this rescaled distribution should
exactly match $\chizero^2\Pinf{\zeta/\xzero^2,\phizero}$, which is
represented as a dashed curve in the figure. We see that for all
values of $\xzero < 0.2$, both distributions are indeed
indistinguishable. The agreement is less good for $\xzero$ close to
unity, which is expected, since such values correspond to a sink
located very close to the reflective boundary, in which case quantum
diffusion is impacted. Nonetheless, although the overall amplitude
does not match for $\xzero \lesssim 1$, the overall shape remains
similar to that of the semi-infinite flat potential. In particular,
$\Pof{\zeta\mid\phizero}$ is always skewed towards positive values and
maximal at a curvature that we numerically determine to saturate at
\begin{equation}
\eval{\dfrac{\zetamode}{\chir^2}}_{\xzero=1} = 7.2 \times 10^{-3}.
\end{equation}
The mode has indeed some dependence on $\xzero$. For small values of
$\xzero$, this dependence can be semi-analytically determined by using
\cref{eq:magicmath}. 
Indeed Refs.~\cite{Blachier:2025tcq, Blachier:2025iwk} showed that the mode of
$\chizero^2\Pinf{\zeta,\phizero}$ occurs at a fixed value
$\zetamodeinf/\chizero^2 = 6.9 \times 10^{-3}$ such that, for the flat
well, one has
\begin{equation}
\eval{\dfrac{\zetamode}{\chir^2}}_{\xzero < 0.2} \simeq 6.9 \times 10^{-3}
\times \xzero^2.
\end{equation}
It vanishes in the limit $\xzero \to 0$, which is indeed compatible with the observed
behaviour in \cref{fig:pzeta}.

\begin{figure}
\begin{center}
  \includegraphics*[width=\onefigw]{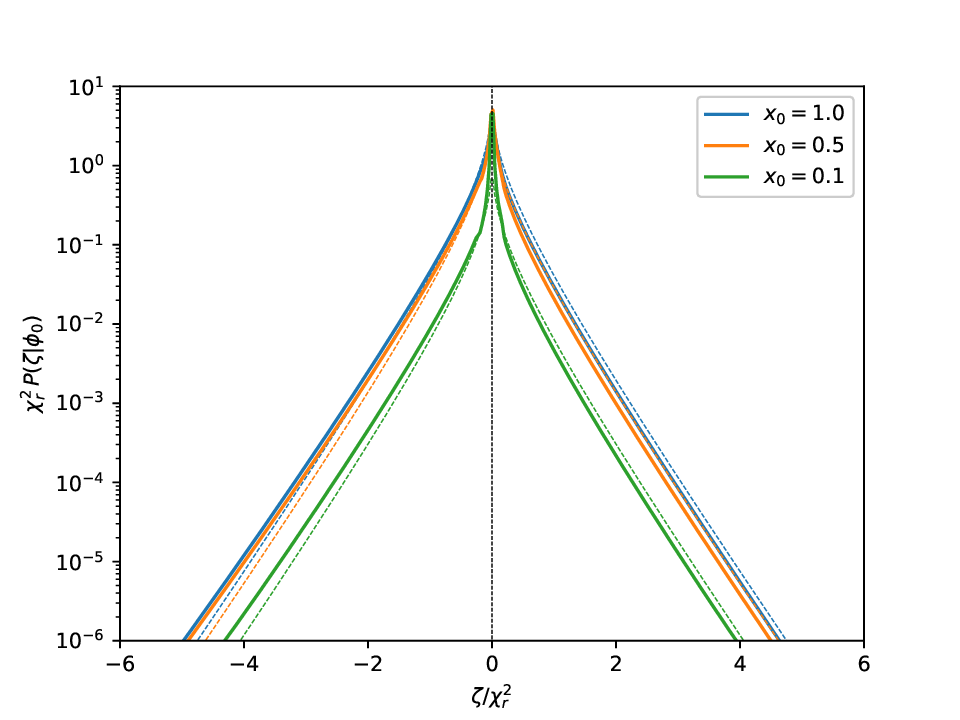}
\caption{Probability distribution for the curvature fluctuations
  $\zeta$ in the quantum well (solid curves), in the domain
  $\abs{\zeta}/\chir^2>1$ and for various values of $\xzero$. The
  dashed curves represent the approximation in \cref{eq:pzetatails}, the tails are (quasi)
  exponential and vary as $\Eoneb{\pi^2 \abs{\zeta}/(4\chir^2)}$.}
\label{fig:pzetatails}
\end{center}
\end{figure}

As discussed above, we expect the tails of $\Pof{\zeta\mid\phizero}$
to differ substantially from those of $\Pinf{\zeta\mid\phizero}$, the
latter scaling as $1/\abs{\zeta}^{3/2}$. From \cref{eq:pzetaexplicit},
in the domain of large fluctuations, $\abs{\zeta}/\chir^2 \gg 1$, the
upper bound of the integral is small and the integration domain only
encompasses regions having $\chirhat \ll 1$. The stochastic dynamics
is then driven by saturated quantum diffusion and the probability
distribution of $\zetahat$ is the rectangle function of
\cref{eq:rectlimit}. From \cref{eq:pzeta}, one obtains
\begin{equation}
\chir^2 \Pof{\zeta\mid\phizero} = -\dfrac{\pi}{2}
\int_0^{\frac{\chir}{\sqrt{2\abs{\zeta}}} \ll 1} \djtheta{2}{\dfrac{\pi}{2} \xzero,e^{-\frac{\pi^2}{2\chirhat^2}}}
\dfrac{\dd{\chirhat}}{\chirhat} \simeq \pi \sin\qty(\dfrac{\pi}{2}
\xzero) \int_0^{\frac{\chir}{\sqrt{2\abs{\zeta}}}}
\dfrac{e^{-\frac{\pi^2}{8\chirhat^2}}}{\chirhat} \dd{\chirhat},
\end{equation}
where we have used the expansion~\eqref{eq:thetaqnull} of the theta
function derivative in $q\to 0^{+}$. Notice the appearance of a
dependence on $\xzero$, which is reintroduced by the marginalisation
over $\Plt{\rv{\Nzero}\mid\phizero}$. The integral in the
  previous expression is an exponential integral
\begin{equation}
\Eone{z} \equiv \int_z^{+\infty} \dfrac{e^{-t}}{t} \dd{t},
\end{equation}
and one has
\begin{equation}
  \begin{aligned}
\lim_{\abs{\zeta}/\chir^2 \gg 1} \chir^2 \Pof{\zeta\mid\phizero} & =
\dfrac{\pi}{2} \sin\qty(\dfrac{\pi}{2}\xzero) \Eone{\dfrac{\pi^2}{4}
  \dfrac{\abs{\zeta}}{\chir^2}} \\ & \simeq \dfrac{2}{\pi}
\sin\qty(\dfrac{\pi}{2}\xzero) \dfrac{\chir^2}{\abs{\zeta}}
e^{-\frac{\pi^2}{4} \frac{\abs{\zeta}}{\chir^2}},
\end{aligned}
\label{eq:pzetatails}
\end{equation}
where the approximation in the second line is obtained by using
$\Eone{z}\simeq e^{-z}/z$ at large $z$. In \cref{fig:pzetatails}, we
have plotted as solid curves the exact probability distribution
$\chir^2\Pof{\zeta\mid\phizero}$ in the large fluctuations domain
$\abs{\zeta}/\chir^2 > 1$, for a few values of $\xzero$. The dashed
curves in this figure are the approximation in \cref{eq:pzetatails} and, up to a
constant shift due to the skewness, they match well the tails of the
exact distribution.

\subsection{Comparison with the forward approach}
\label{sec:comparison}

\begin{figure}
\begin{center}
  \includegraphics*[width=\onefigw]{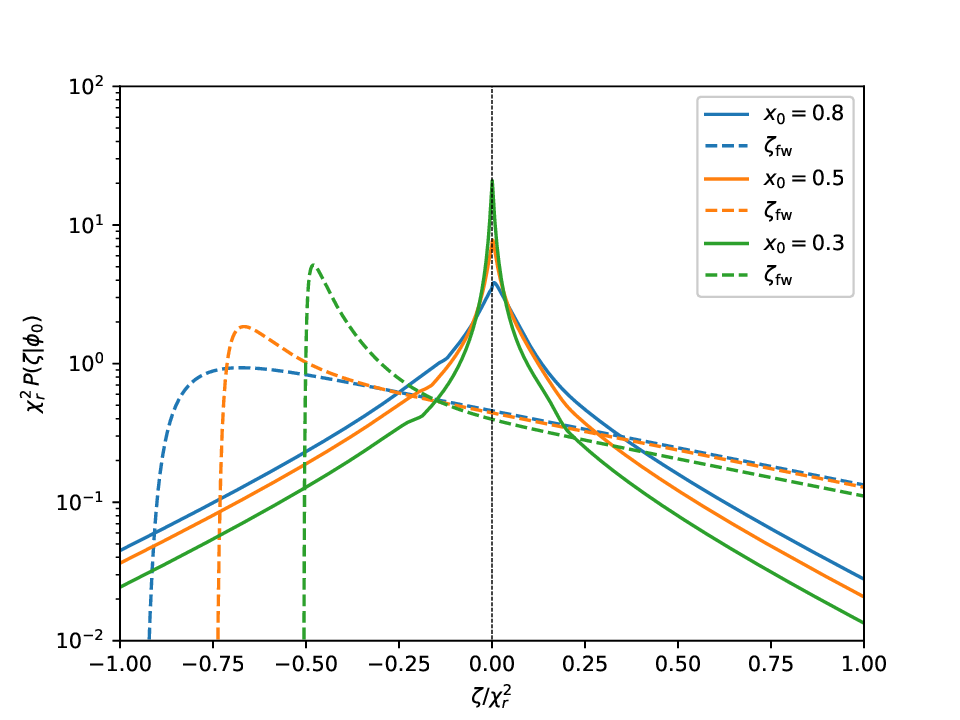}
  \caption{Comparison between the probability distribution for the
    time-reversed curvature fluctuations $\Pof{\zeta\mid\phizero}$ (solid
    curves) and its forward-formalism counterpart
    $\Pof{\zetafw\mid\phizero}$ (dashed curves), for various values of
    $\xzero$. The time-reversed distribution has positive and negative
    exponential tails, whose decay rate is exactly twice that of the positive tail of the forward distribution.}
\label{fig:pzetafw}
\end{center}
\end{figure}

In the forward formalism, the probability distribution for the
curvature fluctuations $\zetafw \equiv \calN - \ev{\calN}$ is given by
\cref{eq:Pzetafwdef,eq:Nflatdef}. The average value here is over the lifetime
realisations only, having a probability distribution given by
\cref{eq:Plt}, i.e., in the flat well, by the first passage times
distribution of \cref{eq:fwPfpt}. This is different from the
time-reversed formalism where $\zeta$ has been determined by the
fluctuations in the number of reverse {\efolds} at given lifetime, see
\cref{eq:zetadef}. One has to evaluate
\begin{equation}
\ev{\rv{\Nzero}} = \int_{0}^{+\infty}\rv{\Nzero} \Plt{\rv{\Nzero}\mid\phizero} \dd{\rv{\Nzero}} =
-\dfrac{\pi^2}{4\chir^2} \int_0^{+\infty} \rv{\Nzero}
\djtheta{2}{\dfrac{\pi}{2}\xzero,e^{-\frac{\pi^2}{2 \chir^2}
    \rv{\Nzero}}} \dd{\rv{\Nzero}}.
\label{eq:meancalNdef}
\end{equation}
Using the series expression of $\djtheta{2}{z,q}$ obtained from
\cref{eq:deftheta2}, this integral can be analytically expanded as an
infinite series and then resumed. A more direct approach is to use the
characteristic function associated with $\Pfpt{\Nqw\mid\phizero}$, which
is the generating functional of all the moments
$\ev{\rv{\Nzero}^n}$. One gets~\cite{Pattison:2017mbe, Ezquiaga:2019ftu, Animali:2022otk}
\begin{equation}
\ev{\rv{\Nzero}} = \chir^2 \xzero \qty(2 - \xzero),
\label{eq:meanlife}
\end{equation}
from which \cref{eq:Pzetafwdef} gives
\begin{equation}
\Pof{\zetafw\mid\phizero} = -\dfrac{\pi}{4 \chir^2}
\djthetab{2}{\dfrac{\pi}{2} \xzero,
  e^{-\frac{\pi^2}{2}\qty(\frac{\zetafw}{\chir^2} + 2\xzero - \xzero^2)}}.
\label{eq:Pzetafw}
\end{equation}
As in the time-reversed picture, the probability distribution $\chir^2
\Pof{\zetafw\mid\phizero}$ is only a functional of $\zetafw/\chir^2$ and
$\xzero$. Notice, however, that \cref{eq:Pzetafw} is only defined for
\begin{equation}
\dfrac{\zetafw}{\chir^2} > - \xzero(2 - \xzero),
\end{equation}
and its analytic continuation is exactly vanishing at that value. This
is related to the very definition of $\zetafw$ that does not allow
negative fluctuations smaller than $-\ev{\rv{\Nzero}}$. In the
positive tail of \cref{eq:Pzetafw}, i.e., for $\zetafw/\chir^2 \gg 1$,
one can again use the expansion~\eqref{eq:thetaqnull} of the theta
function derivative at $q \to 0^{+}$ to get
\begin{equation}
\lim_{\zetafw/\chir^2 \gg 1} \chir^2 \Pof{\zetafw\mid\phizero} = \dfrac{\pi}{2}
\sin\qty(\dfrac{\pi}{2} \xzero) e^{-\frac{\pi^2}{8} \qty(2\xzero -
  \xzero^2)} e^{-\frac{\pi^2}{8}\frac{\zetafw}{\chir^2}}.
\label{eq:Pzetafwtail}
\end{equation}
This expression can be compared with \cref{eq:pzetatails}: the positive
tail of $\Pof{\zetafw\mid\phizero}$ is exponential, but it decays exactly
twice as slowly as that of $\Pof{\zeta\mid\phizero}$. This factor of
two difference in the exponential behaviour of the positive tail was
already encountered in the semi-infinite potential with drift, see
Ref.~\cite{Blachier:2025iwk}. As such, it seems to be a generic
difference between the forward and time-reversed approaches, at least when strong quantum
diffusion is driving the dynamics. As we have shown earlier, the flat
well is indeed either very similar to the semi-infinite flat
potential, or even more ``quantum''. In \cref{fig:pzetafw}, we have
plotted both $\chir^2 \Pof{\zeta\mid\phizero}$ (solid curves) and
$\chir^2 \Pof{\zetafw\mid\phizero}$ (dashed curves) for various values of
$\xzero$. The factor of two difference in the positive tail is readily
visible, as well as the asymmetry between the positive and negative
domains for $\Pof{\zetafw\mid\phizero}$. Notice that for $\xzero \to 0$,
$\Pof{\zetafw\mid\phizero}$ is non-vanishing only for $\zetafw>0$ whereas
$\Pof{\zeta\mid\phizero}$ becomes sharply peaked around $\zeta=0$ with a
shape approaching the semi-infinite limit of \cref{fig:pzeta}. The
positive and negative tails remain exponential, as in
\cref{eq:pzetatails}, while the amplitude of the tails is reduced
according to the factor $\sin\qty(\pi \xzero/2)$.

\section{Conclusion}
\label{sec:concl}

In this work, we have exactly solved time-reversed stochastic
inflation in the finite flat well, i.e., for an exactly flat but
bounded potential having one absorbing and one reflective boundary. At
given lifetime $\rv{\Nzero}$, we have found that quantum diffusion in
the flat well is indistinguishable from the one occurring in a
semi-infinite flat potential~\cite{Blachier:2025tcq}, provided the
well width, measured in units of the typical Brownian excursion, is
large enough: $\chirhat=\rv{\phir}/\qty(G \sqrt{\rv{\Nzero}}) \gg
1$. In the opposite limit of a small well width, $\chirhat \ll 1$, the
system becomes more stochastic: the coarse-grained field $\phi$
explores the entirety of the domain with a uniform distribution in
time, up to erasing any memory of the initial conditions. This regime
has been referred to as saturated quantum diffusion. Therefore, from
the time-reversed point of view, a flat well exhibits more
stochasticity than an unbounded flat potential.

In \cref{sec:curv}, we have derived the probability distribution of
the time-reversed curvature fluctuations $\Pof{\zeta\mid\phizero}$, first
at given lifetime, and then marginalised over all lifetimes, our
results being plotted in \cref{fig:pzeta}. The shape of the
probability distribution in the small fluctuation domain
$\abs{\zeta}/\chir^2 < 1$, where $\chir =\rv{\phir}/G$ is the well
width measured in units of the diffusion coefficient $G=\Hinf/(2\pi)$,
is the same as the one derived in the semi-infinite flat
potential, $\Pinf{\zeta\mid\phizero}$. This is quantified by the scaling relation derived in
\cref{eq:magicmath}. On the contrary, for large fluctuations
$\abs{\zeta}/\chir^2 > 1$, saturated quantum diffusion dominates and
both the negative and positive tails of $\Pof{\zeta\mid\phizero}$ are
found to be (quasi) exponential (see \cref{fig:pzetatails}).

The flat well has been extensively studied in the context of PBHs,
using the standard stochastic formalism, and this allows us to compare
in \cref{sec:comparison} the one-point statistics of the curvature
fluctuations derived from the two approaches, forward and time
reversed. This provides only the second setup in which the two methods
have been compared, alongside the semi-infinite flat potential with a
non-vanishing drift studied in Ref.~\cite{Blachier:2025iwk}. In the
flat well, we recover the same quantitative differences as those
discussed for that previous case: in particular, the positive
exponential tail of the reverse distribution decays exactly twice as
fast as the positive tail of the forward distribution. Other
differences are recovered, such as symmetric tails in the reverse
formalism versus a one-sided distribution in the forward approach. In
the limit of large width, $\chir \to \infty$, only the reverse
distribution converges towards $\Pinf{\zeta\mid\phizero}$, which has
Levy-like tails in $\abs{\zeta}^{3/2}$, the forward distribution
becoming ill-defined.

Let us recap the motivation underlying the time-reversed
stochastic-inflation formalism. Since observers are attached to the
end-of-inflation hypersurface, statistical quantities that have to be
compared with observations should be derived in reverse {\efold}
number~\cite{Sasaki:1998ug}. The time-reversed approach extends these
results to stochastic inflation, where an unperturbed trajectory in
field space does not exist. Do observable predictions differ from the
forward formalism? In the semi-classical regime, the answer is well
known and positive, but not by much, as the differences are given by
slow-roll quantities~\cite{Tada:2016pmk}. In the quantum-diffusion
regime, the present paper suggests that they are indeed
different. Hence, our results may have some important consequences. If
we were to apply these results blindly to primordial black hole
production, we would predict far fewer PBHs, due to exponential tails
decaying twice as fast. Moreover, the exponential tail at negative
$\zeta$ might also have observable consequences, as it predicts an
enhancement of negatively curved regions in a proportion quasi
identical to that of positively curved ones. However, some care has to
be taken before drawing conclusions. As discussed in the introduction,
the time-reversed approach involves an inherent conditioning on the
lifetime, together with a marginalisation over all its realisations,
such that the distribution $\Pof{\zeta\mid\phizero}$ is representative
of the likeliness that our background universe is a quantum
realisation having experienced $\rv{\Nzero}$ total {\efolds} of
quantum diffusion. This may not be the statistics of interest when one
is concerned with the probability of forming PBHs within one of these
realisations. In order to address these questions quantitatively, it
would be more interesting, though quite challenging, to solve
time-reversed stochastic inflation in a system admitting a
semi-classical limit.

\section*{Acknowledgements}

TT thanks the CURL group at University of Louvain for their
hospitality during the visit where this work was initiated.  CA and CR
are supported by the ESA Belgian Federal PRODEX Grants
$\mathrm{N^{\circ}} 4000143201$ and $\mathrm{N^{\circ}}
4000144768$. BB is publishing in the quality of ASPIRANT Research
Fellow of the FNRS.  TT is supported by JSPS KAKENHI Grant Numbers
25K01004, MEXT KAKENHI 23H04515 and 25H01543.  KT is supported by JSPS
Overseas Research Fellowships.

\appendix

\section{Forward solution by Fourier transform}
\label{sec:fwdfourier}

Defining $\PfwFT{K,N}$ such that
\begin{equation}
\Pfw{\phi,N\mid\phizero,\Nzero} = \int_{-\infty}^{+\infty} \PfwFT{K,N}
e^{i K \phi} \dd{K},
\label{eq:FTdef}
\end{equation}
the solutions of \cref{eq:basicheat} read
\begin{equation}
\PfwFT{K,N} = \PfwFT{K,\Nzero} e^{-\frac{G^2 K^2}{2}\left(N-\Nzero\right)}.
\end{equation}
Plugging this solution into \cref{eq:FTdef}, and enforcing that
$\Pfw{\phi,\Nzero\mid\phizero,\Nzero}$ is real, one gets
\begin{equation}
\Pfw{\phi,N\mid\phizero,\Nzero} = \dfrac{1}{\pi} \int_0^{\infty}
e^{-\frac{G^2 K^2}{2}\left(N-\Nzero\right)} \qty[A(K,\Nzero)
  \cos\qty(K \phi) - B\qty(K,\Nzero) \sin\qty(K \phi)]\dd{K},
\label{eq:Pfwexpanded}
\end{equation}
where $A$ and $B$ are the real and imaginary parts of
$\PfwFT{K,\Nzero}$. The boundary conditions of \cref{eq:fwdbic}
require
\begin{equation}
    A \cos\qty(K\phiqw) - B\sin\qty(K\phiqw)  = 0, \qquad A \sin\qty(K \phir) + B\cos\qty(K\phir)  = 0,
\end{equation}
whose solutions are, for $n$ integer,
\begin{equation}
K = K_n \equiv \dfrac{(2n+1) \pi}{2 \rv{\phir}}\,,\qquad
\dfrac{A_n}{B_n} = \tan\qty(K_n \phiqw) = -
\dfrac{1}{\tan\qty(K_n\phir)}\,.
\end{equation}
Solving for $B_n$ and plugging the above solution into
\cref{eq:Pfwexpanded}, one obtains
\begin{equation}
\Pfw{\phi,N\mid\phizero,\Nzero} = \dfrac{1}{\pi} \sum_{n=0}^\infty
C_n \, e^{-\frac{G^2 K_n^2}{2}\left(N-\Nzero\right)} \sin\qty(K_n\rv{\phi}),
\end{equation}
where the yet to be determined coefficients $C_n \equiv
-B_n/\cos\qty(K_n\phiqw)$. They are uniquely set by the initial
condition. Indeed,
\begin{equation}
\Pfw{\phi,N=\Nzero\mid\phizero,\Nzero} = \dirac{\phi-\phizero} = \dfrac{1}{\pi}\sum_{n=0}^\infty C_n \sin\qty(K_n\rv{\phi}),
\end{equation}
which is the sine decomposition of the Dirac distribution. The
coefficients $C_n$ are then given by
\begin{equation}
C_n = \dfrac{2\pi}{\rv{\phir}} \sin\qty(K_n\rv{\phizero}).
\end{equation}
The unique solution of \cref{eq:basicheat,eq:fwdbic} finally reads
\begin{equation}
\Pfw{\phi,N\mid\phizero,\Nzero} = \dfrac{2}{\rv{\phir}} \sum_{n=0}^\infty
e^{-\frac{G^2 K_n^2}{2}\left(N-\Nzero\right)} \sin\qty(K_n\rv{\phizero}) \sin\qty(K_n\rv{\phi}).
\label{eq:Pfwsin}
\end{equation}
This solution can also be expressed in terms of cosine functions
using the trigonometric product-to-sum identities as
\begin{equation}
\Pfw{\phi,N\mid\phizero,\Nzero} = \dfrac{1}{\rv{\phir}} \sum_{n=0}^\infty
e^{-\frac{G^2 K_n^2}{2}\left(N-\Nzero\right)}
\Bqty{\cos\qty[K_n\qty(\rv{\phizero}-\rv{\phi})] - \cos\qty[K_n\qty(\rv{\phizero}+\rv{\phi})]}.
\label{eq:Pfwcos}
\end{equation}
This infinite summation can be reduced to the closed-form expression
of \cref{eq:Pfwtheta} by recognizing the definition of the second
Jacobi theta function written in \cref{eq:deftheta2}.

\section{Time-reversed solution from the Maruyama--Girsanov's method}
\label{sec:revmaruyama}

The Maruyama--Girsanov's theorem quantifies the effect of a change of
probability measure in the stochastic
ensembles~\cite{Maruyama1955ContinuousMP, Girsanov1960}. We consider
$\Bqty{X}$ the ensemble of all possible realisations, up to a time
$t$, of some stochastic process described by
\begin{equation}
\dd{X} =  \Frv(X,t) \dd{t} + \dd{W(t)},
\label{eq:mathprocess}
\end{equation}
where $W(t)$ is a driftless Brownian motion. Let us consider a
friction given by \cref{eq:Frvfpfw}, i.e.,
\begin{equation}
\Frv(x,t) = -\dfrac{\pi}{2} \dfrac{\djthetab{2}{\pi\,
    \dfrac{\xzero-x}{2},e^{-\frac{\pi^2}{2 }\qty(\tzero-t)}} +
  \djthetab{2}{\pi\,\dfrac{\xzero+x}{2},e^{-\frac{\pi^2}{2 }\qty(\tzero-t)}}}{\jthetab{2}{\pi\,
    \dfrac{\xzero-x}{2},e^{-\frac{\pi^2}{2 }\qty(\tzero-t)}} -
  \jthetab{2}{\pi\,\dfrac{\xzero+x}{2},e^{-\frac{\pi^2}{2
      }\qty(\tzero-t)}}}\,,
\label{eq:mathdrift}
\end{equation}
where all physical quantities have been set to unity for
clarity. The Maruyama--Girsanov's theorem implies that, for any
arbitrary functional $h$, one has
\begin{equation}
\Evalb{h\qty(\Bqty{X})} = \Evalb{Z(t)h\qty(\Bqty{W})},
\label{eq:MGtheorem}
\end{equation}
with
\begin{equation}
Z(t) = \exp{\int_0^t \Frv \left[W(u),u\right] \dd{W(u)} -
  \dfrac{1}{2} \int_0^t \Frv^2\left[W(u),u\right] \dd{u}}.
\end{equation}
Therefore, if a simple expression for $Z(t)$ can be found, one
can always express the transition probability distribution of the
process $X$ in terms of the one of a pure Brownian
motion~\cite{Mazzolo2024, Blachier:2025tcq}.

Let us consider the following primitive of the drift term~\eqref{eq:mathdrift}
\begin{equation}
f(x,t) = \ln\Bqty{\jthetab{2}{\pi\dfrac{\xzero-x}{2},e^{-\frac{\pi^2}{2 }\qty(\tzero-t)}} -
  \jthetab{2}{\pi\dfrac{\xzero+x}{2},e^{-\frac{\pi^2}{2 }\qty(\tzero-t)}} }.
\label{eq:fprim}
\end{equation}
Its It\^o differential reads~\cite{Doeblin1940, Ito1944}
\begin{equation}
\dd{f\qty[W(u),u]} = \eval{\left( \pdv{f}{t} + \dfrac{1}{2}
\pdv[2]{f}{x} \right)}_{W(u),u}\dd{u} + \eval{\pdv{f}{x}}_{W(u),u} \dd{W(u)}.
\end{equation}
By construction, the factor multiplying $\dd{W}$ is the drift
$\Frv(W,u)$. Moreover, using the identity
\begin{equation}
q \pdv{\jtheta{2}{z,q}}{q} = -\dfrac{1}{4} \pdv[2]{\jtheta{2}{z,q}}{z},
\end{equation}
one gets, from \cref{eq:fprim}, the simple relation
\begin{equation}
\pdv{f}{t} + \dfrac{1}{2}\pdv[2]{f}{x} = -\dfrac{1}{2} \Frv^2(x,t),
\end{equation}
where all terms involving the second derivatives of $\jtheta{2}{z,q}$
cancel. The It\^o differential simplifies to
\begin{equation}
  \dd{f\qty[W(u),u]} = \Frv(W,u) \dd{W} - \dfrac{1}{2}{\Frv^2(W,u)}
  \dd{u},  
\end{equation}
which allows us to calculate $Z(t)$ explicitly
\begin{equation}
Z(t) = \dfrac{e^{f\qty[W(t),t]}}{e^{f\qty[W(0),0]}} = \dfrac{\jthetab{2}{\pi\dfrac{\xzero-W(t)}{2},e^{-\frac{\pi^2}{2 }\qty(\tzero-t)}} -
  \jthetab{2}{\pi\dfrac{\xzero+W(t)}{2},e^{-\frac{\pi^2}{2
      }\qty(\tzero-t)}}}{\jthetab{2}{\pi\dfrac{\xzero-W(0)}{2},e^{-\frac{\pi^2}{2 }\tzero}} -
  \jthetab{2}{\pi\dfrac{\xzero+W(0)}{2},e^{-\frac{\pi^2}{2
      }\tzero}}}\,.
\end{equation}
The solution for the driftless Brownian motion starting at $\wzero$, in presence of one
absorbing and one reflective boundary, has already been derived in
\cref{sec:fwdfourier} and it is given in \cref{eq:Pfwtheta}, i.e.,
\begin{equation}
\Pfw{w,t\mid\wzero,0} = \dfrac{1}{2} \Bqty{\jthetab{2}{\pi\dfrac{\wzero-w}{2},e^{-\frac{\pi^2}{2
    }t}} - \jthetab{2}{\pi\dfrac{\wzero+w}{2},e^{-\frac{\pi^2}{2
      }t}}}.
\end{equation}
From \cref{eq:MGtheorem}, one gets the probability
distribution for the process described by \cref{eq:mathprocess}
\begin{equation}
  \begin{aligned}
\Pof{x,t\mid\wzero,0} &= \dfrac{\jthetab{2}{\pi\dfrac{\xzero-x}{2},e^{-\frac{\pi^2}{2 }\qty(\tzero-t)}} -
  \jthetab{2}{\pi\dfrac{\xzero+x}{2},e^{-\frac{\pi^2}{2
      }\qty(\tzero-t)}}}{\jtheta{2}{\pi\dfrac{\xzero-\wzero}{2},e^{-\frac{\pi^2}{2 }\tzero}} -
  \jtheta{2}{\pi\dfrac{\xzero+\wzero}{2},e^{-\frac{\pi^2}{2
      }\tzero}}} \\ & \times \dfrac{\jtheta{2}{\pi\dfrac{\wzero-x}{2},e^{-\frac{\pi^2}{2
    }t}} - \jtheta{2}{\pi\dfrac{\wzero+x}{2},e^{-\frac{\pi^2}{2
      }t}}}{2}\,.
\end{aligned}
\end{equation}
For the time-reversed solution we are interested in, the initial state has
to be chosen on the absorbing boundary and taking the limit
$\wzero\to0$ in the previous expression gives the transition
probability for the reverse process
\begin{equation}
  \Prv{x,t} =
  \dfrac{\djtheta{2}{\dfrac{\pi}{2} x,e^{-\frac{\pi^2}{2}t}}}{2
    \djtheta{2}{\dfrac{\pi}{2}\xzero,e^{-\frac{\pi^2}{2}\tzero}}}
    \Bqty{\jthetab{2}{\pi\dfrac{\xzero-x}{2},e^{-\frac{\pi^2}{2 }\qty(\tzero-t)}} -
  \jthetab{2}{\pi\dfrac{\xzero+x}{2},e^{-\frac{\pi^2}{2
      }\qty(\tzero-t)}}},
\end{equation}
where use of $\jtheta{2}{z,q} = \jtheta{2}{-z,q}$ has been made.

\section{Normalisation of the curvature fluctuation distribution at given lifetime}
\label{sec:normalisation}

We prove here that, within the time-reversed formalism, the
distribution of $\zetahat=\zeta/\rv{\Nzero}$, the curvature
fluctuation in units of the lifetime, at given lifetime, is always
normalised to unity. From \cref{eq:P:zetahat:prop:Pbar}, recall that
\begin{equation}
    \Pof{\zetahat\mid\phizero,\rv{\Nzero}} = \int_{0}^{1} \dd{x} \Pof{x, \zetahat\mid\phizero,\rv{\Nzero}}
    \qty[\heaviside{\ev{\tau} - \zetahat} - \heaviside{\ev{\tau} -
    \zetahat - 1}],
\end{equation}
where $\ev{\tau}$ depends on $x, \phizero$ and $\rv{\Nzero}$. Integrating the above expression
over $\zetahat$ and commuting the two integrals, one obtains
\begin{equation}
\begin{aligned}
    \int_{-\infty}^{+\infty} \dd{\zetahat} \Pof{\zetahat\mid\phizero,\rv{\Nzero}} &=  \int_{0}^{1} \dd{x} \int_{-\infty}^{+\infty} \dd{\zetahat} 
    \Pof{x, \zetahat\mid\phizero,\rv{\Nzero}}\\ 
    &\times \qty[\heaviside{\ev{\tau} - \zetahat} - \heaviside{\ev{\tau} -
    \zetahat - 1}].
    \end{aligned}
\end{equation}
Now that the two integrals are exchanged, one can perform the change
of variable, $s = \ev{\tau} - \zetahat$, for each $x$, since
$\ev{\tau}$ depends on $x$. The two Heaviside functions impose that
$0<s<1$ so that, after commuting back again the order of the two integrals,
\begin{equation}
    \int_{-\infty}^{+\infty} \dd{\zetahat}
    \Pof{\zetahat\mid\phizero,\rv{\Nzero}} = \int_{0}^{1} \dd{s}
    \int_{0}^{1} \dd{x} \Prv{x, s \mid\phizero,\rv{\Nzero}}.
\label{eq:norm}
\end{equation}
From an operator point of view, see
e.g. Ref.~\cite{Sarkka_Solin_2019}, the Fokker-Planck equation has the
mass conservation property~\cite{Perthame2015}. It can be
straightforwardly extended to its time-reversed version so that the
integral over all field values of the transition probability $\Prv{x,
  s \mid\phizero,\rv{\Nzero}}$ is ensured to be a constant,
independent on $s$. Evaluating it in $s=0$ fixes the constant to $1$
since, in that limit, it converges to a Dirac distribution. The
right-hand side of \cref{eq:norm} therefore evaluates to $1$. Hence,
the distribution of $\zetahat$ is always normalised to unity. Had we
used $\zeta$ instead of $\zetahat$, the integral would have evaluated
to $\rv{\Nzero}$. This is consistent with \cref{eq:proba_conserv} that
merely states probability conservation.

\bibliography{Bibliography}
\bibliographystyle{JHEP}

\end{document}